\documentclass[usenatbib,useAMS]{mnras}

\usepackage[utf8]{inputenc}
\usepackage[T1]{fontenc}
\usepackage{ae,aecompl}
\usepackage{newtxtext,newtxmath}

\usepackage{graphicx}	
\usepackage{amsmath}	
\usepackage{amsfonts}
\usepackage{multicol}   
\usepackage{bm}		    
\usepackage{pdflscape}	
\usepackage{acro}
\usepackage[table,xcdraw]{xcolor}
\usepackage{tikz}
\usepackage{algorithm} 
\usepackage{algpseudocode}
\usepackage[capitalise]{cleveref}
\usepackage[normalem]{ulem}

\definecolor{bittersweet}{rgb}{1.0, 0.44, 0.37}

\definecolor{amber}{rgb}{1.0,0.49,0.0}
\definecolor{purple}{RGB}{128, 0, 128}

\DeclareAcronym{DM}{
  short = DM,
  long  = dark matter
}

\DeclareAcronym{NN}{
  short = NN,
  long  = neural network
}

\title[The scatter in the galaxy--halo connection]{The scatter in the galaxy--halo connection: a machine learning analysis}

\author[R. Stiskalek et al.]
{Richard Stiskalek,$^{1}$\thanks{\href{mailto:richard.stiskalek@protonmail.com}{richard.stiskalek@protonmail.com}} 
Deaglan J. Bartlett,$^{2}$\thanks{\href{mailto:deaglan.bartlett@physics.ox.ac.uk}{deaglan.bartlett@physics.ox.ac.uk}}
Harry Desmond,$^{2,3,4}$\thanks{\href{mailto:harry.desmond@port.ac.uk}{harry.desmond@port.ac.uk}}
and Dhayaa Anbajagane$^{5,6}$\thanks{\href{mailto:dhayaa@uchicago.edu}{dhayaa@uchicago.edu}}
\\ 
$^{1}$Universit\"{a}ts-Sternwarte, Ludwig-Maximilians-Universit\"{a}t M\"{u}nchen, Scheinerstr. 1, 81679 München, Germany\\
$^{2}$Astrophysics, University of Oxford, Denys Wilkinson Building, Keble Road, Oxford, OX1 3RH, UK\\
$^{3}$McWilliams Center for Cosmology, Department of Physics, Carnegie Mellon University, 5000 Forbes Ave, Pittsburgh, PA 15213, USA\\
$^4$Institute of Cosmology \& Gravitation, University of Portsmouth, Dennis Sciama Building, Portsmouth, PO1 3FX, UK\\
$^5$Department of Astronomy and Astrophysics, University of Chicago, 5640 S. Ellis Ave, Chicago, IL 60637, USA\\
$^6$Kavli Institute for Cosmological Physics, University of Chicago, 5640 S. Ellis Ave, Chicago, IL 60637, USA
}

\date{Accepted 2022 June 7. Received 2022 May 30; in original form 2022 February 28}

\pubyear{2021}

\begin{document}
\label{firstpage}
\pagerange{\pageref{firstpage}--\pageref{lastpage}}
\maketitle

\begin{abstract}
We apply machine learning, a powerful method for uncovering complex correlations in high-dimensional data, to the galaxy--halo connection of cosmological hydrodynamical simulations. The mapping between galaxy and halo variables is stochastic in the absence of perfect information, but conventional machine learning models are deterministic and hence cannot capture its intrinsic scatter. To overcome this limitation, we design an ensemble of neural networks with a Gaussian loss function that predict probability distributions, allowing us to model statistical uncertainties in the galaxy--halo connection as well as its best-fit trends. We extract a number of galaxy and halo variables from the Horizon-AGN and IllustrisTNG100-1 simulations and quantify the extent to which knowledge of some subset of one enables prediction of the other. This allows us to identify the key features of the galaxy--halo connection and investigate the origin of its scatter in various projections. We find that while halo properties beyond mass account for up to $50$ per cent of the scatter in the halo-to-stellar mass relation, the prediction of stellar half-mass radius or total gas mass is not substantially improved by adding further halo properties. We also use these results to investigate semi-analytic models for galaxy size in the two simulations, finding that assumptions relating galaxy size to halo size or spin are not successful.
\end{abstract}


\section{Introduction}\label{sec:Introduction}

The dynamics of the Universe are governed by \ac{DM}, yet our knowledge of this in galaxies comes entirely from the photometry and spectroscopy of baryons, with five times lower average density. Relating the behaviour of baryons and \ac{DM} is a key task of astrophysics. \ac{DM} may be usefully described by the halo model, in which \ac{DM} clumps into gravitationally bound quasi-spherical structures called haloes, which in turn host cold baryons in the form of galaxies. In this regime, relating baryons and \ac{DM} is tantamount to discovering the correlations between galaxies and haloes: the galaxy--halo connection. This would be the first step towards uncovering the physical processes responsible for the baryon--dark matter relation in a bottom-up approach to galaxy formation.

With \ac{DM} being unobservable, the galaxy--halo connection cannot be inferred directly. Instead we are forced to do this either empirically---by studying the properties of galaxies such as their kinematics which are determined by \ac{DM}---or by postulating scaling relations between galaxy and halo variables and calculating how they would affect observables: examples include halo abundance matching~\citep{Kravtsov_2004,Behroozi_2010}, the halo occupation distribution~\citep{HOD} (and embellishments thereof;~\citealt{Paranjape_HOD, Decorated_HOD, Hadzhiyska_HOD}), the angular momentum partition model of~\cite{MMW_1998} and the rich field of semi-analytic modelling (see~\citealt{Baugh} for a review). A third approach is to infer the galaxy--halo connection post-hoc from cosmological hydrodynamical simulations, which model numerically the coevolution of baryons and \ac{DM} from their initial conditions with a mix of physical prescriptions and semi-empirical subgrid models and feedback schemes. While each simulation implicitly produces its own galaxy--halo connection, this is often far from clear from the raw output of the simulation, nor is it obvious how this comes about or what the implications are for other types of modelling.

We focus here on the third approach. As the correlations between galaxy and halo properties in a given simulation are nonlinear and have unknown functional forms a priori, the best approach for reconstructing them is to leverage \emph{machine learning} (ML) whereby a computer can rapidly test a huge range of possibilities.Indeed, machine-learning the galaxy--halo connection is already becoming a cottage industry, particularly in supervised settings where algorithms are trained on data sets generated by empirical, semi-analytic or simulation-based models. For example, \cite{vonMarttens_2021} explored a variety of supervised models, such as Support Vector Machine or Random Forests. Among the tree-based models, the Random Forest has been used extensively to study the galaxy--halo connection~\citep{Man_2019,Delgado_2021,Lovell_2021} along with e.g. Gradient Boosted Decision Trees~\citep{Chen_2021}. Similarly, the use of \ac{NN}-based models
is becoming increasingly popular, with fully-connected networks~\citep{Moster_GalaxyNet,Shao_2021}, Graph Neural Networks~\citep{Villanueva_Domingo_2021} and Convolutional Neural Networks~\citep{de_los_Rios_2021} all having recently been brought to bear. The advantage of this approach is that---if successful---it enables a faithful reconstruction of the galaxy--halo connection of a simulation, as well as aiding its interpretability. The latter may be done either with a feature importance analysis, or by identifying analytic expressions that are successful at describing the connection, e.g. through symbolic regression~\citep{Delgado_2021}. Recent work has also demonstrated the power in utilising dedicated ML interpretibility tools---such as Google DeepDream~\citep{Ntampaka2019MLDeepDream} and the Shapley values~\citep{Machado2021SHAP, Anbajagane_2022}---in determining key galaxy or halo properties. From a practical point of view, ML approaches are also useful for painting galaxies onto the haloes produced in N-body, \ac{DM} only simulations---or haloes onto observed galaxies---at low computational expense compared to full hydrodynamical simulation. The disadvantage is that the results are tied to a particular simulation with no guarantee of accuracy should that simulation fail to describe the real Universe. Indeed, recent results have indicated that ML algorithms trained on one simulation will not in general be successful at reproducing others~\citep{Villaescusa_Navarro_2021}.

A key limitation of many previous ML analyses is that the scatter in predicted variables tends to be underestimated \citep{Kamdar_2016,Agarwal_2018} because mock catalogues are generated using a deterministic ML algorithm such that a given input always implies the same output. Such algorithms do not reflect the nature of the galaxy--halo connection, which is inherently stochastic in the absence of complete information. The main aim of this work is to construct and showcase a ML algorithm which returns not only best-fit predictions for galaxy properties given halo properties (or vice versa) but also the \emph{statistical uncertainty} on those predictions. This latter goal has been pursued by only a minor subset of existing ML works \citep[e.g.][]{Ramanah2020MLUncertainties, Ramanah2021MLUncertainties, Ho2021MLUncertainties, Eisert2022ErgoML}. It is crucial to capture the scatter in the galaxy--halo connection, and this will allow us to quantify the extent to which this scatter correlates with other galaxy/halo properties such that it can be reduced or removed when including those properties in the feature set. With this algorithm in hand, we ask which galaxy properties can be predicted from their hosts' (sub)halo properties and vice versa, and which features are the most important for making these predictions. To test the robustness of these results to modelling assumptions we use two different cosmological hydrodynamical simulations, Horizon-AGN and IllustrisTNG, as training sets. Along the way we will determine the extent to which some popular semi-empirical galaxy formation models---halo abundance matching and DM-baryon angular momentum partition---are successful at describing the simulations' outputs.

In~\Cref{sec:Data sets} we introduce the simulations and describe the simulated galaxy and halo properties we use as features and targets for our ML algorithm. Our ML architecture is outlined in~\Cref{sec:methodology}, where we also discuss how we reduce the impact of correlations in the input features and feature selection. We present the correlations of the simulated galaxy--halo connections in~\Cref{sec:Results}, and their implications for semi-empirical models in~\Cref{sec:test_sam}. \Cref{sec:Discussion} discusses the results and suggests further work, and \Cref{sec:Conclusion} concludes.

\section{Simulated data sets}\label{sec:Data sets}

We describe the simulations we use in~\Cref{sec:Cosmological simulations}, and the definitions of the halo/galaxy properties and how we compute them in~\Cref{sec:halo_galaxy_defs}.

\subsection{Cosmological simulations}\label{sec:Cosmological simulations}

\subsubsection{Horizon-AGN}\label{sec:HAGN_simulation}

For one of our training sets, we consider the galaxies and their corresponding haloes in the $(100 {\rm \, Mpc}/h)^3$ cosmological hydrodynamical simulation Horizon-AGN\footnote{\url{http://www.horizon-simulation.org/about.html}} (henceforth H-AGN)~\citep{Dubois_2014}. This adopts a standard $\Lambda$CDM cosmology consistent with WMAP-7~\citep{Komatsu_2011} and therefore has total matter density $\Omega_{\rm m} = 0.272$, dark energy density $\Omega_{\rm \Lambda} = 0.728$, amplitude of the matter power spectrum $\sigma_8 = 0.81$, baryon density $\Omega_{\rm b} = 0.045$, Hubble constant $H_0 = 70.4 \, {\rm \, km \, s^{-1} \, Mpc^{-1}}$, and power spectrum slope $n_{\rm s} = 0.967$.

Using $1024^3$ \ac{DM} particles to give a \ac{DM} mass resolution of $M_{\rm DM, \, res} = 8 \times 10^7 {\rm \, M_{\sun}}$, the simulation is run with the Adaptive Mesh Refinement code \textsc{ramses}~\citep{Teyssier_2002}, with cells refined whenever the mass in that cell surpasses 8 times the initial mass resolution, up to an effective physical resolution of $1{\rm \, kpc}$. Stars can form when the hydrogen number density exceeds $n_{\rm H} \geq 0.1 {\rm \, H \, cm^{-3}}$, yielding a stellar mass resolution of $M_{\rm \star, \, res} \approx 2 \times 10^6 {\rm \, M_{\sun}}$, and this process follows a Schmidt law with a $1$ per cent efficiency~\citep{Rasera_2006}. Baryonic effects such as background UV heating, gas cooling (including a contribution from metals), feedback from stellar winds and type Ia and type II supernovae assuming a Salpeter initial mass function (IMF)~\citep{Dubois_2008,taysun2012a} are also incorporated in the simulation. Following~\cite{Dubois_2013} we split the gas associated with each galaxy into the following four phases by its temperature $T$ and hydrogen number density $n_{\rm H}$
\begin{itemize}
    \item Star forming: $n_{\rm H} \geq 0.1~\mathrm{H~cm}^{-3}$, which is the star formation threshold in H-AGN,
    \item Hot diffuse: $n_{\rm H} < 0.1~\mathrm{H~cm}^{-3}$ and $T\geq 10^5~\mathrm{K}$,
    \item Cold collimated: $100 \: n_{\rm avg} \leq n_{\rm H} < 0.1~\mathrm{H~cm}^{-3}$ and $T < 10^5~\mathrm{K}$, where $n_{\rm avg}$ is the mean hydrogen number density of the Universe,
    \item Cold diffuse: $n_{\rm H} < 100~n_{\rm avg}$ and $T < 10^5~\mathrm{K}$.
\end{itemize}

Using the \textsc{adaptaHOP} structure finder~\citep{Aubert_2004,Tweed_2009}, galaxies (haloes) consist of a minimum of 50 (100) particles, and are identified when the stellar (\ac{DM}) smoothed density field calculated using the 20 nearest neighbours exceeds 178 times the mean total matter density~\citep{Gunn_1972}. A shrinking sphere approach~\citep{Power_2003} finds the position of the densest particle, which is taken to be the centre of the structure. Galaxies are then paired to haloes by considering each halo in turn from the most to least massive, and choosing the most massive unassigned galaxy which lies within 10 per cent of the virial radius, $R_\mathrm{vir}$, of the halo as its partner~\citep{Chisari_2017,Bartlett_OffsetBH_2021}. All galaxy and subhalo properties for H-AGN are computed using all available particles within $R_{\rm vir}$, and this includes a minor fraction of gravitationally unbound material as well.

We enforce the following threshold on our resulting catalogue: halo mass $M_\mathrm{h} > 10^{10} M_\odot$, corresponding to $125$ particles, stellar mass $M_\star > 10^{8.3} M_\odot$, corresponding to $100$ particles, and the galaxy average star forming rate over the last $300~\mathrm{Myrs}$ be greater than $0$ since some star-forming galaxies may be assigned a star formation rate (SFR) of $0$ due to the true SFR being below the resolution limit.

\subsubsection{TNG}\label{sec:TNG-100}

Our second training set consists of the galaxies and haloes from the TNG100-1 simulation\footnote{\url{https://www.tng-project.org/}}~\citep{Marinacci_2018,Naiman_2018,Nelson_2018,Pillepich_2018_results,Springel_2018} (henceforth TNG). Like H-AGN, TNG adopts a $\Lambda$CDM cosmology, however the parameters are those from the 2015 analysis of \textit{Planck}~\citep{Planck_Parameters_2016}: $\Omega_{\rm m} = 0.3089$, $\Omega_{\rm \Lambda}=0.6911$, $\Omega_{\rm b}=0.0486$, $\sigma_8=0.8159$, $H_0 = 67.74 \, {\rm \, km \, s^{-1} \, Mpc^{-1}}$ and $n_{\rm s} = 0.9667$.

This hydrodynamical simulation is run using the moving-mesh code \textsc{arepo}~\citep{Springel_2010} with a volume of $(75 {\rm Mpc}/h)^3$, which is $\sim50$ per cent of H-AGN's volume. The~\citet{Springel_2003} model of star formation and pressurisation of the multi-phase interstellar medium is used, such that stars form stochastically according to the Kennicutt-Schmidt relation with a Chabrier IMF~\citep{Chabrier_2003} when the gas density exceeds $n_{\rm H} \approx 0.1 {\rm \, cm^{-3}}$. Numerous baryonic effects are included such as feedback from active galactic nuclei (AGNs), galactic winds, metal cooling, and ideal magneto-hydrodynamics; for full details see~\citet{Weinberger_2017} and~\citet{Pillepich_2018_methods}.

The Friends-of-Friends (FoF) algorithm is run on the \ac{DM} particles, with a linking length $b=0.2$, to determine the main haloes, and other particle species are associated with the halo of their closest \ac{DM} particle. The \textsc{subfind} algorithm~\citep{Springel_2001,Dolag_2009} is then used on all particle species to identify substructures, and hence galaxies. The most massive subhalo within a main halo is defined to be the central subhalo, and the others are classified as satellite subhaloes.

The \textsc{subfind} algorithm, when identifying subhaloes within FoF haloes, also performs an unbinding step to remove any material that is not gravitationally bound to the subhalo. All subhalo properties for TNG are computed using only the bound particles. Matching H-AGN, we define our sample so that halo mass $M_{\rm h} > 10^{10} M_\odot$, corresponding to $1300$ particles, stellar mass $M_\star > 10^{8.3} M_\odot$, corresponding to $140$ particles, and the instantaneous SFR in the subhalo is larger than $0$.

\begin{figure*}
    \centering
    \includegraphics[width=\textwidth]{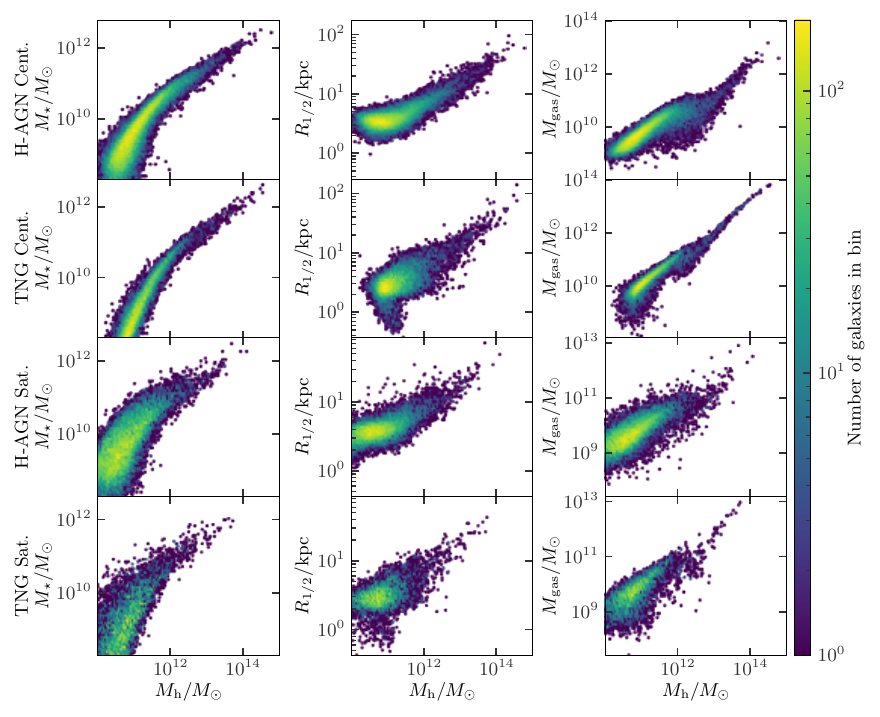}
    \caption{H-AGN and TNG correlations of $M_\mathrm{h}$ with $M_\star$ (left column), $R_{1/2}$ (central column) and $M_\mathrm{gas}$ (right column) for central and satellite galaxies separately. Rows are ordered as H-AGN centrals, TNG centrals, H-AGN satellites and TNG satellites. The colour scale indicates logarithmic number counts.}
    \label{fig:sim_data}
\end{figure*}

\subsection{Halo and galaxy property definitions}\label{sec:halo_galaxy_defs}

We describe here the full set of halo and galaxy properties used in this work, including differences in definitions used by each simulation and the public catalogues utilised for some of the properties.

In general, a key difference between simulation properties is that the H-AGN properties are computed using all particles (bound or unbound) within a spherical aperture cut, whereas the TNG properties use only the gravitationally bound particles without any such aperture cut. This arises from differences in the subhalo finders used by each simulation.

\subsubsection{Subhalo/galaxy mass and radius}\label{sec:Mass_Radius}

The subhalo radius is utilised to define the subhalo mass, in addition to being an input property in our method, so we begin by describing it. For H-AGN, this is obtained by binning the DM particles in concentric ellipsoidal shells, and determining the outermost shell for which the virial theorem is satisfied to within $20$ per cent. The virial radius, $R_{\rm vir}$, is defined to be the geometric mean of the three axis radii of this ellipsoid. If no such shell exists, then the virial radius is defined to be the largest radius within which the mean density exceeds that expected from a spherical top-hat model. For parent haloes this procedure yields a near one-to-one relation between the radius and virial mass. This matches the typical virial radius definition:
\begin{equation}\label{eqn:virmass}
    \Bar{\rho}(<R_{\rm vir}) \equiv \frac{M(<R_{\rm vir})}{\frac{4}{3}\pi R_{\rm vir}^3} = \Delta_{\rm vir}\rho_c,
\end{equation}
where $\Bar{\rho}$ is the average enclosed density, $M(<r)$ is the enclosed mass, $\Delta_{\rm vir}$ is the overdensity as defined in~\citet{BryanNorman1998}, and $\rho_c$ is the critical density of the Universe at the present epoch. For subhaloes the radius specified in~\cref{eqn:virmass} acts as an upper limit, with a significant fraction having smaller sizes. Therefore, H-AGN subhaloes' sizes may be an informative quantity even when their masses are already included in the feature set. 

In TNG parent haloes the virial radius is defined to obey~\cref{eqn:virmass}. For TNG, we define the subhalo radius as the radial distance to the furthest gravitationally bound DM particle in the subhalo particle set determined by the \textsc{subfind} algorithm. Henceforth, we refer to all subhalo radii as $R_h$. The subhalo masses for both H-AGN and TNG are then simply the total mass of all available DM particles within $r < R_{\rm h}$.
In TNG subhaloes we define the virial mass of the host halo as $M_{\rm vir}^{\rm P} \equiv M(<R_{\rm vir})$. On the other hand, for H-AGN subhaloes we define $M_{\rm vir}^{\rm P}$ to be the virial DM mass of their parent haloes.

The stellar/gas mass of a subhalo ($M_\star,\,M_{\rm gas}$) in each simulation is defined in a way analogous to how their total subhalo mass is defined above. For H-AGN, it is the total mass of all stars/gas within $R_{\rm vir}$, whereas for TNG it is the total mass of all stars/gas bound to the subhalo. In TNG, the stellar half-mass radius, $R_{\rm 1/2}$ is then defined as the radius containing half of the total stellar mass. In H-AGN we project the stellar particles onto three orthogonal planes, and compute the two-dimensional radius for each plane which contains half the stellar mass. $R_{1/2}$ is then the geometric mean of these radii.

\subsubsection{\Acl{DM} subhalo concentration}\label{sec:halo_concentration}

The density profiles of subhaloes can be parameterised according to the Navarro-Frenk-White (NFW) model~\citep{Navarro_1996},
\begin{equation}
\label{eq:nfw_profile}
	\rho \left( r \right) = \frac{\rho_0}{\left( r / r_{\rm s} \right) \left( 1 + r / r_{\rm s} \right)^{2}},
\end{equation}
where $\rho_0$ and $r_{\rm s}$ are model parameters denoting the characteristic density and the scale radius, respectively. Once we fit the model to the profile of a given subhalo, we define the concentration as
\begin{equation}
    c_{\rm h} \equiv \frac{R_{\rm h}}{r_{\rm s}}.
\end{equation}
Note that this is analogous to the standard definition but using $R_{\rm h}$ rather than $R_{\rm vir}$. For TNG, $r_s$ is obtained by fitting the \emph{enclosed mass} NFW profile rather than the density profile (as is done for H-AGN) where the two profiles are related by a simple radial, volume integral.

For central subhaloes, we compared our TNG catalogue with the TNG mass--concentration relation from~\citet*{Anbajagane_2022}, who performed a traditional density profile fit to determine $r_s$, and verified that the difference between the two $r_s$--mass relations is less than five per cent. For central subhaloes, we also confirmed that defining $c_{\rm h}$ using $R_{\rm h}$ instead of a traditional spherical overdensity radius results in a $c_{\rm h} - M_{\rm h}$ relation that still captures the relevant mass-dependent features induced by baryonic effects \citep*[see ``wiggle'' features in][Figure 3]{Anbajagane_2022}.

\subsubsection{\Acl{DM} subhalo and galaxy shapes}

We parameterise the shape of a subhalo/galaxy by first calculating the inertia tensor
\begin{equation}
    \label{eq:inertia tensor}
    I^a_{ij} = \frac{1}{M_{\rm a}} \sum_n m_{\rm a}^{\left( n \right)} x_i^{\left( n \right)}x_j^{\left( n \right)},
\end{equation}
where the sum runs over the particles of type ``a'' (where ``a'' is \ac{DM} when computing \ac{DM} subhalo shapes, and stars when computing galaxy shapes), each of mass $m_a^{\left( n \right)}$ and at position $\bm{x}^{\left( n \right)} = \left(x^{\left(n\right)}_1, x^{\left( n \right)}_2, x^{\left( n \right)}_3 \right)^\intercal$. The total mass of the particle set is denoted $M_a$. We diagonalise $I^a_{ij}$ to find the eigenvalues $\mathcal{A}_a^2 \geq \mathcal{B}_a^2 \geq \mathcal{C}_a^2$, where $\mathcal{A}_a,\mathcal{B}_a,\mathcal{C}_a > 0$~\citep{Chisari_2015}. We use the ratios of the eigenvalues (i.e. the principal axes) to obtain two shapes,
\begin{equation}
    s_a \equiv \mathcal{C}_a/\mathcal{A}_a, \quad q_a \equiv \mathcal{B}_a/\mathcal{A}_a,
\end{equation}
conforming to the notation of denoting halo properties with subscript ``h'' and galaxy properties with subscript ``g''.

As with \ac{DM} concentration, we compare the TNG central subhalo shapes with the TNG catalogues of~\citet*{Anbajagane_2022}, who computed \ac{DM} shapes by selecting particles within a chosen spherical aperture. Given we instead use the full \ac{DM} subhalo (as determined by \textsc{subfind}) without any spherical cut, the shapes we find are $5-10$ per cent more elliptical, i.e. the ratios $s_{\rm h}$ and $q_{\rm h}$ are lower. This is an expected feature due to the difference in particle selection~\citep{Zemp2011Shapes}. However, the functional form of the shape-mass relation is the
same regardless of whether such an aperture cut is used; in particular, the mass-dependent slopes decrease linearly (in log-mass) towards higher masses \citep*[see][Figure 5]{Anbajagane_2022}.

\subsubsection{Spins}

The spins of the \ac{DM}, $\lambda_{\rm h}$, and the stars, $\lambda_g$, are computed as,
\begin{equation}
    \lambda_{\{h,\,g\}} = \frac{|\Vec{J}_{\{h,\,g\}}|}{\sqrt{2}M_{\rm h}R_{\rm h}V_{\rm h}}
\end{equation}
where $\Vec{J}_{\{h,\,g\}}$ is the sum of the angular momenta of the particles, either \ac{DM} or stars, within the subhalo radius as defined above. The normalisation is the self-similar expectation for the total angular momentum in a halo, defined using its mass and radius. The velocity scale is $V_{\rm h} = \sqrt{GM_{\rm h}/R_{\rm h}}$, with the mass and radius definitions described above. This definition of spin is analogous to that of~\citet{Bullock_2001}, but with different mass/radius conventions.

\subsubsection{Other galaxy properties}\label{sec:other_gal_props}

We use the time-averaged SFR, defined as
\begin{equation}
    {\rm SFR} = \frac{M_\star(\Delta t)}{\Delta t},
\end{equation}
where $M_\star(\Delta t)$ is the total stellar mass formed over the last $\Delta t$ years. For the time-scale, we use $300$ Myr for H-AGN and $200$ Myr in TNG, with the difference purely set by the available catalogues. For TNG, we obtain this quantity from the publicly released catalogues~\citep{Donnari_2019, Pillepich_2019}.

The mass-weighted mean stellar metallicity, $Z$, is computed using stellar particles within twice the stellar half-mass radius of the subhalo centre. We also consider the galaxy rest-frame photometry, namely the absolute magnitudes $M_u, M_g, M_r, M_i$ computed using the \textit{u, g, r, i} band filters of the Sloan Digital Sky Survey~\citep{Kaviraj_2017,Laigle_2019}. The magnitude of a galaxy is obtained by summing up the luminosities of individual stellar particles.

For TNG, we also use $M_{\rm HI}$, which is the total neutral hydrogen mass in the subhalo, and this is once again obtained from public catalogues~\citep{Diemer_2018,Diemer_2019}.

\vspace{3mm}

\noindent We show the correlations of $M_h$ with $M_\star$, $R_{1/2}$ and $M_\text{gas}$ for central and satellite galaxies in H-AGN and TNG in~\cref{fig:sim_data}. The full set of galaxy and halo properties is summarised in Table~\ref{tab:properties}.

\begin{table*}
    \centering
    \begin{tabular}{ccc}
        Property & H-AGN & TNG100 \\
        \hline
        \hline
        $R_{\rm h}$ & Radius within which \ac{DM} (sub)halo is both bound and virialised & Distance from halo centre to furthest bound \ac{DM} particle\\
        $M_{\rm h}$ & Total mass of \ac{DM} within $R_{\rm h}$ & Total mass of \ac{DM} bound to (sub)halo\\
        $c_{\rm h}$ & NFW subhalo concentration ($R_{\rm h}/r_{\rm s}$) & Same as H-AGN$^\dagger$\\
        $M_{\rm vir}^P$ & Virialised DM mass of subhalo's parent halo & Spherical overdensity total virial mass of subhalo's parent halo\\
        $\lambda_{\rm h}$ & Bullock spin of \ac{DM} particles & Same as H-AGN$^\dagger$\\
        $s_{\rm h},\,q_{\rm h}$ & Ratios of principal axes of \ac{DM} mass ellipsoid & Same as H-AGN$^\dagger$\\
        \hline
        $M_{\star}$ & Total mass of stars within $R_{\rm h}$ & Same as H-AGN$^\dagger$\\
        $M_{\rm gas}$ & Total mass of gas within $R_{\rm h}$ & Same as H-AGN$^\dagger$\\
        $R_{\rm 1/2}$ & Radius containing half the total stellar mass & Same as H-AGN$^\dagger$\\
        $\rm SFR$ & Time-averaged star formation rate over the last $300$ Myr & Time-averaged star formation rate over the last $200$ Myr\\
        $\rm Z$ & Mass-weighted metallicity of stars & Same as H-AGN$^\dagger$\\
        $\rm M_{u, g, r, i}$ & Absolute magnitude of galaxy in different bands & Same as H-AGN$^\dagger$\\
        $\rm M_{\rm HI}$ & N/A & Total mass of neutral hydrogen bound to (sub)halo\\
        $\lambda_{\rm g}$ & Bullock spin of stellar particles & Same as H-AGN$^\dagger$\\
        $s_{\rm g},\,q_{\rm g}$ & Ratios of principal axes of stellar mass ellipsoid & Same as H-AGN$^\dagger$\\
        \hline
    \end{tabular}
    \caption{Summary of the different (sub)halo/galaxy properties of each simulation used in this analysis, including key differences. The definitions are discussed in detail in~\Cref{sec:halo_galaxy_defs}. \\[5pt]
    $^\dagger$ Note that while H-AGN and TNG calculate properties using similar methods/definitions, the particle set that is operated on differs noticeably between the two. H-AGN uses all available particles (both gravitationally bound and unbound) within a spherical aperture cut, whereas TNG only uses bound particles and without any spherical aperture cut.}
    \label{tab:properties}
\end{table*}

\section{Methodology}
\label{sec:methodology}

In~\Cref{sec:sim_scatter} we describe how we calculate bivariate scatters in the simulations to set the stage for full multivariate scatter calculated by our \acp{NN}, then in~\Cref{sec:preprocess} discuss the data preprocessing, in~\Cref{sec:NNarchitecture} the architecture of our \ac{NN} model and in~\Cref{sec:NN_prediction} the interpretation of its predictions. Lastly in~\Cref{sec:decision_trees} and~\Cref{sec:feature_selection} we describe the Extra-Trees model and feature selection, respectively.

\subsection{Quantifying the bivariate scatter and its uncertainty}
\label{sec:sim_scatter}

A key property of the galaxy--halo connection is its intrinsic scatter, i.e. the variation across the population of galaxy/halo quantities conditional on one or more others. Developing a framework to model this---along with its uncertainty---is a key goal of our work. A simple benchmark for the scatter may be obtained by specialising to the correlation between just one galaxy property and one halo property. In this case the scatter provides an upper limit on how well one can determine one property from the other (i.e. in the absence of any further information) and can be calculated straightforwardly from the simulated quantities. If the residuals are correlated with a third property then including that in the relation will reduce the scatter and hence allow a more precise prediction of the target variable.

Consider the correlation between two quantities $X$ and $Y$, where we wish to predict $Y$ given $X$. To quantify the scatter in the $X-Y$ relation and its uncertainty, we first sort the data by $X$ into $n_X$ bins between the minimum and maximum values of $X$ (typically we choose bins of width $0.15{\, \rm dex}$ when $X$ is the halo mass). We assume that in each bin $B_i$, $Y$ is Gaussian distributed about a mean $\mu_i$ with standard deviation $\sigma_i$. Assuming each galaxy is independent, the likelihood for the observed $Y$ in bins $\{B_i\}$ given these parameters is
\begin{equation}
	 \mathcal{L} \left( \{ Y_g \}_i | \mu_i, \sigma_i \right)
	 = \prod_{g \in B_i} \mathcal{N}\left(Y_g | \mu_i, \sigma^2_i\right),
\end{equation}
where $Y_g$ is the value of $Y$ for galaxy $g$, $\{ Y_g \}_i \equiv \{ Y_g : g \in B_i \}$, and $\mathcal{N}( Y_g | \mu_i, \sigma^2_i )$ is a Gaussian of mean $\mu_{i}$ and variance $\sigma_i^2$. Given a prior, $p \left( \mu_i, \sigma_i \right)$, on $\mu_i$ and $\sigma_i$, we use Bayes' theorem to obtain the posterior distributions of $\mu_i$ and $\sigma_i$,
\begin{equation}
	p \left(\mu_i, \sigma_i^2| \{ Y_g \}_i \right) 
    \propto \mathcal{L} \left(\{ Y_g \}_i|\mu_i, \sigma_i \right) p \left( \mu_i, \sigma_i \right) ,
\end{equation}	
which we sample using the Markov chain Monte Carlo sampler \textsc{EPSIE}\footnote{\href{https://cdcapano.github.io/epsie/latest/index.html}{www.cdcapano.github.io/epsie}}~\citep{epsie}. Within each bin $B_i$ we assume a uniform prior on $\mu_i$ between the minimum and maximum values of  $\{Y_{\rm g}\}_i$ and a uniform prior on $\sigma_i$ between $0$ and the difference between the maximum and minimum values of $\{Y_{\rm g}\}_i$.

We run this procedure separately for each $B_i$ and obtain samples from the two-dimensional posterior distributions. We find the one-dimensional marginalised posterior for each $\sigma_i$,
\begin{equation}
    p \left(\sigma_i| \{ Y_g \}_i \right) = \int p \left(\mu_i, \sigma_i| \{ Y_g \}_i \right) {\rm d}\mu_i,
\end{equation}
by projecting the posterior samples onto the $\sigma_i$ axis. We quote the scatter in the $X$--$Y$ relation for bin $B_i$ as the median of $p \left(\sigma_i| \{ Y_g \}_i \right)$ and its upper and lower uncertainties as the absolute difference between the median and the 84\textsuperscript{th} and 16\textsuperscript{th} percentiles of $p \left(\sigma_i| \{ Y_g \}_i \right)$, respectively. This procedure is designed to penalise bins with few samples, since although the quoted scatter for those bins may be small, it will have a large uncertainty.

Note that the above models the scatter as being in the $Y$ direction, which is appropriate for predicting $Y$ given $X$. This is affected by the slope of the $Y-X$ relation, so a more suitable measure of intrinsic scatter in general may be that orthogonal to the mean relation. However, since we are interested in the prediction of $Y$ given $X$, we do not consider this further. We will use these results to quantify the extent to which including additional variables reduces the scatter in the galaxy--halo connection.

\subsection{Machine learning preprocessing}\label{sec:preprocess}

We split the data into into training ($80$ per cent) and test ($20$ per cent) sets using a stratified shuffle split, ensuring that the split preserves the same fraction of samples as in the original set when binned by the target property. We reserve $20$ per cent of the training set for validation, which is used for either hyperparameter selection or deciding whether to continue training.
The test set remains unseen during training and is only used to evaluate the performance of a trained model. All results are evaluated on the test set.

To account for the fact that most galaxy and halo properties span large dynamic ranges we work with their logarithms, except in case of magnitudes, principle axes ratios and star metallicity. We apply the ``standard'' scaling to transform the features, enforcing a mean of $0$ and variance of $1$. We also transform the features into an orthogonal basis (as outlined below) because most machine learning algorithms display marginally superior performance on uncorrelated features. All transformations are fitted on the training set.

The properties of galaxies and haloes quoted in the simulation catalogues form a mutually correlated basis. We map them to a linearly uncorrelated ``orthogonal'' basis using principal component analysis (PCA)~\citep{pearson_PCA,hotelling_PCA} such that the components maximise the variance of the data. Dimensionality reduction could then be done by projecting into a subspace containing only components with high variance, although we do not do so here because it is not necessary due to a small number of variables and because it discards information.

\subsection{Neural network architecture}\label{sec:NNarchitecture}

To model the galaxy--halo connection we use an ensemble of feed-forward \acp{NN} with a Gaussian loss function. A feed-forward NN defines a mapping $\mathcal{F}: \mathbb{R}^m \rightarrow \mathbb{R}^n$ with $m$ input nodes (input layer) and $n$ output nodes (output layer), possibly with hidden layers in between. The simplest fully connected NN with no hidden layers is a single-layer perceptron which connects the input layer $m$-dimensional vector $x_i$ directly to the output $n$-dimensional vector $y_i$ such that
\begin{equation}
	y_i = f\left(\sum_{j=1}^{n} W_{ij} x_j + b_i\right),
\end{equation}
where $W_{ij}$ is a $n\times m$-dimensional matrix of weights and $b_i$ is a bias vector, both of which are trainable parameters, and $f(\cdot)$ is an activation function. For a network with matrix multiplication layers (as outlined above) to learn a non-linear mapping, the activation function $f(\cdot)$ must be non-linear. Common choices of $f(\cdot)$ are the sigmoid function or hyperbolic tangent. If a network contains hidden layers then the vector $y_i$ is  used as input to the next layer. It has been shown that a NN is capable of approximating any Borel-measurable function to any desired accuracy, thus acting as a universal approximator, provided sufficient hidden nodes are used~\citep{Hornik_universal_approximator}.

We choose a scaled exponential linear unit (SELU) activation function~\citep{Klambauer_SELU}, defined as
\begin{equation}
    f(x) = \lambda
    \begin{cases}
        x &\quad \mathrm{if}~ x \geq 0\\
        \alpha \left(e^x - 1\right) &\quad \mathrm{if}~ x < 0,
    \end{cases}
\end{equation}
where $\lambda = 1.05$ and $\alpha = 1.67$ are chosen to induce self-normalisation of the network, which preserves zero mean and unit variance between consecutive layers if the initial weights are drawn to satisfy these constraints on average. Networks with the SELU activation are robust to perturbations, can be efficiently used to train deep feed-forward networks and avoid the problem of vanishing or diverging gradients~\citep{Klambauer_SELU}.

The network parameters are trained by minimising a loss function, $L$, which assesses the closeness of the prediction $\mathcal{F}(\bm{x})$ to $\bm{y}$. We train the network on known tuples of features and targets $\left\{\bm{x}, \bm{y}\right\}$ (\emph{supervised learning}). They are fitted using backpropagation which computes the gradient of $L$ with respect to the trainable parameters~\citep{Rumelhart_backprop}. With the gradient available the optimal set of parameters can be found via e.g. a gradient descent algorithm.

During training, the gradient descent optimiser searches the parameter space of network weights and bias vectors until a global minimum is found. If the loss function is interpreted as the negative log-likelihood, then the minimum corresponds to the maximum-likelihood estimate. However, in practice the parameter space has a complex structure and, therefore, the optimiser can converge to a local minimum instead.

A common loss function used to infer the optimal networks parameters is the mean square error (MSE). MSE can be viewed as optimising the mean network prediction, without assessing the prediction uncertainty. To account for the uncertainty~\cite{Nix_mean_variance} suggest to use as a loss function the negative log-likelihood of a Gaussian distribution: 
\begin{equation}
    \label{eq:Gaussian loss function}
    L\left(\left\{y\right\} | \left\{\bm{x}\right\}, \bm{\bm{\theta}}\right) 
    =  \sum_k\left[ \ln \sigma_{\bm{\theta}}^{2}(\bm{x}^k) + \frac{\left(y^k - \mathcal{F}_{\bm{\theta}}(\bm{x}^k)\right)^2}{2\sigma_{\bm{\theta}}^{2}(\bm{x}^k)}\right] + \mathrm{const.},
\end{equation}
where $\bm{\theta}$ denotes the network's parameters, $\mathcal{F}_{\bm{\theta}} (\bm{x}^k)$ and $\sigma_{\bm{\theta}}(\bm{x}^k)$ are the network's mean prediction and standard deviation, respectively, and $k$ indexes the tuples of features and targets. In practice, this is achieved by having the network's final layer output two values: both the mean and standard deviation of the Gaussian. We apply a so-called soft-plus transformation to the standard deviation to enforce positivity:
\begin{equation}
    \sigma_{\bm{\theta}} \to \ln \left( 1 + e^{\sigma_{\bm{\theta}}} \right).
\end{equation}

We use an ensemble of deep \acp{NN} with the aforementioned Gaussian loss function trained with adversarial examples to obtain robust uncertainty estimates~\citep{Lakshminarayanan_deep_ensembles}.
We obtain the adversarial samples $\bm{x}^\prime$ following the fast gradient sign method by perturbing the original samples $\bm{x}$ in the direction of the highest loss $L$:
\begin{equation}\label{eq:fast_gradient}
    \bm{x}^\prime
    =
    \bm{x} + \epsilon~\mathrm{sign}\left(\bm{\nabla}_{\bm x} L\left(\left\{y\right\} | \left\{\bm{x}\right\}, \bm{\bm{\theta}}\right)\right),
\end{equation}
where $\epsilon$ is the constant adversarial step size~\citep{Goodfellow_adversarial_examples}. The target $y$ of each sample remains unchanged under the adversarial attack.
Ensembles of base models typically improve the predictive power~\citep{Dietterich_ensembles}. Building an ensemble typically requires a randomisation scheme to decrease the correlation between the individual base learners. However, the training of a \ac{NN} is a stochastic process which already provides the necessary randomisation to decorrelate the base learners, thus avoiding the need for e.g. bootstrap aggregating as is done with Random Forests (described in~\cref{sec:decision_trees}).

For our \ac{NN} architecture, we use a wide and deep neural network (WDNN) suggested in~\cite{Cheng_wide_deep} and used to model the galaxy--halo connection in~\cite{Moster_GalaxyNet}. A WDNN consists of a wide part which connects the input layer directly to the output layer and a multi-layer perceptron part with several hidden layers. The outputs of both parts are added. An advantage of WDNN is that it learns the linear relations through the wide part, while non-linear relation can be learnt through the deep part of the network. This improves the network's generalisation to unseen samples without loss of accuracy and also improves the network's convergence.

We implement the ensemble of WDNN networks trained with adversarial examples in \textsc{TensorFlow}\footnote{\url{https://www.tensorflow.org}}~\citep{tensorflow2015-whitepaper} and \textsc{TensorFlow Probability}\footnote{\url{https://www.tensorflow.org/probability/}}~\citep{TensorFlow_probability}. The \ac{NN} hyperparameters include the number, type and size of hidden layers, activation function, choice of the gradient descent algorithm and its own hyperparameters, and training procedure. Our procedure for selecting these is as follows.

The network's input layer is normalising: it removes the training samples' mean and rescales them to unit variance. The deep part of the base WDNN network contains $16\times 16\times 16\times 16 \times 8$ hidden nodes whose weights we initialise with the LeCun normal initialiser~\citep{lecun-98b} with the SELU activation function.  We choose the ensemble size to be $250$, the adversarial loss function's weight to be $0.5$,the adversarial step size to be $\epsilon=0.001$, and use the Adam stochastic optimiser~\citep{Adam_Kingma} to find the optimal network parameters with a decaying learning schedule. We set the initial rate to $0.004$ and enforce it to decay by a factor of $1.3$ every $3000$ iterations.
During training the training set is split into batches of size $1000$. These batches are used to update the networks' weights. A single iteration through all batches defines one epoch and we train the networks until the loss function evaluated on the validation set has not lowered for at least $100$ epochs. The samples in the validation set are \emph{not} used to update the network's weights.
We make our \ac{NN} implementation publicly available at \url{https://github.com/Richard-Sti/scatternn}.

\subsection{Quantifying the neural networks' predictions}
\label{sec:NN_prediction}

The {\it q}\textsuperscript{th} base network of the ensemble returns a Gaussian distribution on a target property $y$ centred at $\mu_q$ with standard deviation $\sigma_q$. We average the individual Gaussian distributions, yielding a distribution that can be summarised with mean and variance, respectively
\begin{subequations}\label{eq:Combine NN}
    \begin{align}
        \mu &= \frac{1}{N} \sum_{q=1}^{N} \mu_q,\\
        \sigma^2 &= \frac{1}{N} \sum_{q=1}^{N}\sigma_q^2 + \left(\mu_q - \mu\right)^2,
    \end{align}
\end{subequations}
where $N$ is the number of base networks. We consider the mean of this distribution as the ``best-fit prediction'' and standard deviation as the scatter on the predicted halo/galaxy property.

The ensemble of \acp{NN} models the scatter in galaxy (halo) quantities conditioned on the halo (galaxy) quantities it has been trained on, which may be of any number. In~\cref{sec:sim_scatter} we described our prescription for calculating the bivariate simulation scatter, and its uncertainty, between two properties. This scatter then serves as an upper limit if a third property is added to the feature set, potentially allowing for a more precise prediction.

We compare the bivariate scatter to the \acp{NN}' scatter. If the \acp{NN} are aware of only the property on which the bivariate scatter is conditioned, then we we expect the two to be approximately equal as both are fitting a Gaussian conditioned on the same quantity. However, if the networks learn from other informative properties as well, then the networks' predictions will be not only more precise, i.e. with lower scatter, but this scatter will also be less uncertain.

\subsection{Decision tree ensemble}
\label{sec:decision_trees}

To check the results of our network against a completely different architecture, we also consider non-parametric regression by means of a decision tree ensemble. A decision tree is built from the ``root'' node, which contains the entire training set, by recursively splitting the training set into more homogeneous ``child'' nodes. This is achieved by minimising some loss function $L$ at each split. Typically, in regression the split is chosen such that it minimises the weighted variance of the samples assigned to the left and right child nodes. If the $i$th sample contains several features then the split is chosen along the one that minimises $L$. The splitting is done recursively until a termination condition is met, e.g. the node's variance is zero, it contains only a single sample or a maximum tree depth is reached. Prediction on unseen samples is performed by propagating an input feature vector to a terminal leaf and, in regression, returning the mean value of all training sample targets assigned to the particular terminal leaf.

If the maximum depth of a decision tree is not capped it can ``memorise'' the entire training set. Because of this over-fitting such unregulated decision trees do not generalise well to unseen samples. To account for this the tree can either be ``pruned'' by removing branches with only a small increase in homogeneity, or randomised. A popular algorithm is the Random Forest (RF) which builds an ensemble of decision trees~\citep{Breiman_2001_RandomForest}. RF injects randomness into the ensemble by training the trees on bootstrap aggregated samples and randomly selecting the number of features that are considered for an optimal split. The RF's prediction is the average of the values returned by its constituent trees.

Another ensemble algorithm choice is the Extremely Randomised Trees (Extra-Trees)~\citep{Geurts_2006_ExtraTrees}. Unlike RF, Extra-Trees trees are all grown on the same original sample and randomness is injected in the splitting. After randomly selecting $k$ out of $m$ features, each feature's threshold is \emph{randomly} selected, instead of calculating the optimal threshold, with the best threshold being used to split the node. By picking the thresholds at random the computational cost of training is significantly reduced, while also reducing the model's variance at the cost of a small increase in bias. Moreover, in most applications---including ours---Extra-Trees is as good a predictor as RF, and we will therefore use it as an alternative to our NN model. While Extra-Trees and the other aforementioned tree-based models yield purely deterministic predictions, there exist decision tree models that model the statistical uncertainty~\citep[e.g.][]{BART, NGBoost}. However, we do not consider these here.

We use the \textsc{scikit-learn}\footnote{\url{https://scikit-learn.org/}}~\citep{scikit-learn} implementation of Extra-Trees \textsc{ExtraTreesRegressor}. We find the optimal hyperparameters via a grid search, choosing from $\textsc{n\_estimators}=64, 128$, $\textsc{min\_samples\_split}=8, 16, 32$, $\textsc{max\_samples} = 0.5, 1$ and keeping the default values for the remaining hyperparameters.

\subsection{Feature selection and importance}
\label{sec:feature_selection}

To assess the goodness-of-fit of the mean NN ensemble or Extra-Trees predictions on the test set, we use the $R^2$ coefficient of determination, defined as
\begin{equation}
    R^2
    = 1 - \frac{\sum_{i} \left(y_i - \mathcal{F}(\bm{x}_i)\right)^2}{\sum_{i} \left(y_i - \hat{y}\right)^2},
\end{equation}
where the sums run over the (test set) samples, $\mathcal{F}(\bm{x}_i)$ is the predicted value,  and $\hat{y}$ is the mean target value on the test set. $R^2$ is upper bounded by 1, which signifies perfect predictions (zero residuals). Lower values indicate worse predictions. If $R^2 = 0$ the model only learns to predict the mean value $\hat{y}$ and $R^2 < 0$ signifies a non-informative model whose predictions are imbalanced with respect to the true values on average.

We select important features for predicting a given target by first considering an original set of $N$ features. Then, we train Extra-Trees on all $N$ features individually and select the one that yields the highest $R^2$ on the test set. We keep this feature and, one at a time, add additional properties from the remaining features. We keep whichever new property yields the maximum $R^2$ on the test set when used in conjunction with the features already included. Repeating this process, we end up with an ordered list of features and their $R^2$ increments. We consider as informative all features whose $R^2$ increment is more than $0.005$.

We use the Spearman's rank correlation coefficient $\rho_{\rm s}$ to determine whether the residuals are correlated with any other property that was not used in training. The Spearman coefficient is defined as the Pearson correlation coefficient between rank variables of two vectors, which are formed by replacing the vector's values by their increasing ranks. The Pearson coefficient of two vectors is given by their covariance divided by the product of their standard deviations. The Spearman coefficient is bounded to $\left[-1, 1\right]$, such that $+1$ signifies a correlated monotonic relation and $-1$ an anti-correlated monotonic relation.

Having found an ordered list of features that maximises the $R^2$ increments as described above, we calculate the Spearman correlation coefficient between residuals of a model trained on the first $n$ features and the next feature to be added. This will allow us to further interpret the significance of each added feature.

\section{Correlations of the simulated galaxy--halo connections}
\label{sec:Results}

\subsection{Halo to galaxy mapping}
\label{sec:halo_to_galaxy}

We use the $z=0$ halo properties as features and for simplicity only the following galaxy properties as targets: stellar mass $M_\star$, total gas mass $M_{\rm gas}$ and stellar-half mass radius $R_{1/2}$ (see~\Cref{tab:properties}). Furthermore, in TNG we will also consider learning the mapping from halo properties to the neutral hydrogen mass $M_{\rm HI}$.

For each galaxy target, we begin by training Extra-Trees to determine the important halo features separately in H-AGN and TNG central and satellite galaxies, as described in~\cref{sec:feature_selection}. Since the scatter in the galaxy--halo connection is not crucial for this part, we opt for Extra-Trees rather than the NN ensemble as it provides similar goodness-of-fit of the best-fit predictions while being computationally cheaper. Having determined the important features, we train the NN ensemble, first on the most important feature only, which we find to be $M_{\rm h}$ in all cases, and then we add the remaining important features for each target. We calculate the $R^2$ scores of the mean predictions on the test set, which we then compare to the equivalent inference using Extra-Trees, finding that the two typically agree to within $1$ per cent for well-predicted properties and to within $\sim$ 10 per cent if $R^2 \lessapprox 0.5$. Where there is disagreement, the \ac{NN} may outperform Extra-Trees.
This gives us confidence to employ the \ac{NN} ensemble to measure the scatter of the relations. We compare the \ac{NN} ensemble scatter to the bivariate scatter around the mean in the relation $\log Y | \log M_{\rm h}$, where $Y$ represents one of the galaxy target properties. We illustrate the result by plotting mean and standard deviation of the predicted scatter of test set samples binned by $\log M_{\rm h}$.

We now proceed to describe our results for each separate galaxy target that we consider.

\begin{figure*}
    \centering
    \includegraphics[width=\textwidth]{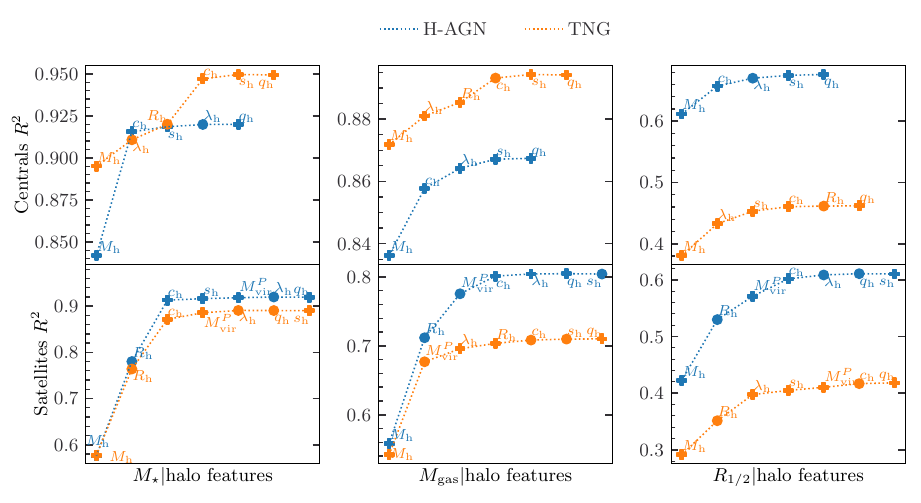}
    \caption{
    Test set $R^2$ scores from Extra-Trees as a function of the cumulative number of features used to predict $M_\star$ (\emph{left} column),  $M_\mathrm{gas}$ (\emph{central} column) and $R_{1/2}$ (\emph{right} column) in H-AGN and TNG central (\emph{top} row) and satellite (\emph{bottom} row) galaxies. The results from the NN ensemble are the same to within $1$ per cent. Features are added in the order that maximises the $R^2$ increments as described in~\cref{sec:feature_selection}. The labels show the halo feature being added to the set of features to the left of it. We indicate whether the residuals of a model trained on the previous set of features have a positive or negative Spearman correlation coefficient with the added feature by marking the point with a ``plus'' or ``circle'', respectively.}
    \label{fig:halo2gal_R2}
\end{figure*}

\subsubsection{Predicting the stellar mass}\label{sec:mstar_pred}

We begin by predicting the stellar mass $M_\star$. Using our feature selection routine outlined in~\Cref{sec:feature_selection} with Extra-Trees we find the maximum possible
$R^2$ to be $0.92$/$0.95$ ($0.91$/$0.90$) in H-AGN/TNG centrals (satellites), which are calculated using all informative halo properties that yield $\Delta R^2 > 0.005$. The left column of~\cref{fig:halo2gal_R2} shows how the $R^2$ depends on the cumulative set of training features. We now describe the informative features in decreasing order of importance.

In H-AGN central galaxies the only important feature beyond $M_{\rm h}$ is $c_{\rm h}$. As indicated by the ``plus'' marker symbol in~\cref{fig:halo2gal_R2}, the residuals of a model trained on $M_{\rm h}$ are positively correlated with $c_{\rm h}$. In TNG centrals the secondary important features are $\lambda_{\rm h}$, $R_{\rm h}$ and $c_{\rm h}$. We note that after including $\lambda_{\rm h}$ and $R_{\rm h}$ the increment in $R^2$ due to $c_{\rm h}$ is larger than the previous two increments. So $c_{\rm h}$ provides noticeable gains in constraining $M_\star$, but only once we have provided information on both $\lambda_{\rm h}$ and $R_{\rm h}$; it is not as useful a feature if $M_{\rm h}$ is the only other input provided to the model. We have not explored the detailed cause for this behaviour, but note that it appears only in the TNG data set and not in H-AGN.

In TNG $M_{\rm h}$ by itself is a better predictor of $M_\star$ compared to H-AGN, corresponding to the tighter $M_\star - M_{\rm h}$ relation shown in~\cref{fig:sim_data}. For satellite galaxies, in H-AGN the important features beyond $M_{\rm h}$ are $R_{\rm h}$ and $c_{\rm h}$. Similarly, in TNG satellites the other relevant features are $R_{\rm h}$, $c_{\rm h}$, $M_{\rm vir}^{P}$ and $\lambda_{\rm h}$, the importances of which are similarly shown in~\cref{fig:halo2gal_R2}. In both simulations we find that after accounting for the $M_{\rm h}$ dependence the halo size $R_{\rm h}$ is anti-correlated with $M_\star$ (``circle'' marker symbol).

\begin{figure*}
    \centering
    \includegraphics[width=\textwidth]{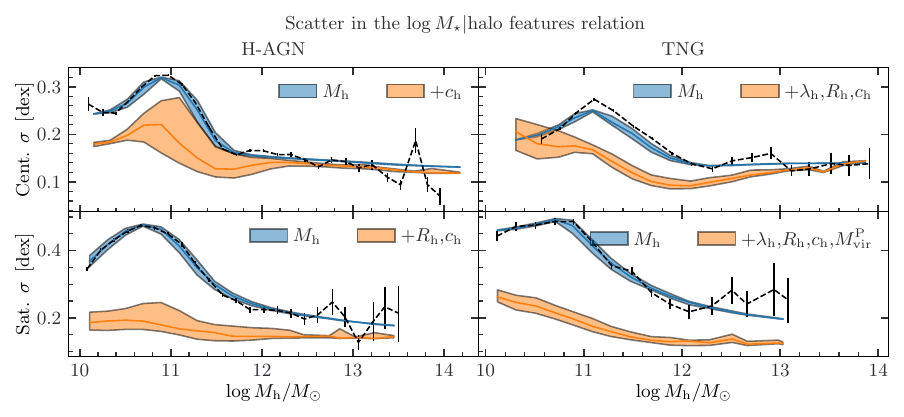}
    \caption{The \acp{NN}' predicted scatter in $\log M_\star$ binned by $\log M_\mathrm{h}$ in H-AGN (\emph{left} column) and TNG (\emph{right} column) for central (\emph{top} row) and satellite (\emph{bottom} row) galaxies. The legend indicates cumulative halo features used to train the networks, with the bands enclosing the $16^{\rm th}$ to $84^{\rm th}$ percentile of predictions for the scatter for galaxies in a given bin of $\log M_{\rm h}$. The black dashed line is the $\sigma_{\log M_\star | \log M_\mathrm{h}}$ bivariate scatter calculated directly from the simulated points, with a $1$ $\sigma$ error bar obtain by fitting Gaussian distributions within fixed-size $\log M_{\rm h}$ bins as described in~\Cref{sec:sim_scatter}. The networks correctly recover the bivariate scatter if trained on $M_{\rm h}$ only (blue band). The orange band includes all other features deemed informative for the prediction of $M_\star$, resulting in a significantly lower uncertainty in $M_\star$ and indicating the extent to which these features correlate with residuals of the $M_\star-M_\mathrm{h}$ relation.
    }
    \label{fig:NNscatter_Mstar}
\end{figure*}

We now move to a study of the scatter in the galaxy--halo correlations. We indicate the bivariate scatter $\sigma_{\log M_\star | \log M_{\rm h}}$ as the black dashed line in~\cref{fig:NNscatter_Mstar} for both H-AGN and and TNG. The simulations agree that above $M_\mathrm{h} \approx 10^{12} M_\odot$ the scatter is $\sim0.15$ and $\sim0.2~\mathrm{dex}$ for centrals and satellites, respectively. Below this mass the scatter starts growing until $M_\mathrm{h} \approx 10^{11} M_\odot$, where it artificially peaks due to the lower limit on $M_\star$ that we impose in both simulations. In H-AGN centrals the scatter peaks $\sim 0.05~\mathrm{dex}$ higher than in TNG centrals, reflecting the tighter $M_\star - M_{\rm vir}$ relation shown in~\cref{fig:sim_data}. The rise in vertical scatter is largely a consequence of the steepening of the $M_\star-M_\text{h}$ relation at the faint end (see~\cref{fig:sim_data}).

We train the NN ensemble model on the features found above to be important. In all cases, we verify that the best-fit predictions and scatter are consistent with the true values on the test set by inspecting the fraction of objects with true target properties lying within the scatter interval. We plot the NN predicted scatter as the blue and orange bands in~\cref{fig:NNscatter_Mstar}, again binned by halo mass for direct comparison with the bivariate scatter. In all cases, training the ensemble on $M_\mathrm{h}$ alone correctly recovers the scatter in the simulations' $\log M_\star | \log M_\mathrm{h}$ relations, showing that the ensemble is able to capture the nonlinear $M_\star-M_\text{h}$ relation.

In H-AGN centrals the addition of $c_{\rm h}$ reduces the scatter below $M_{\rm h}\approx 10^{12} M_\odot$, whereas in TNG centrals the correlation of $M_\star-M_\text{h}$ residuals with $\lambda_{\rm h}$, $R_{\rm h}$ and $c_{\rm h}$ explains up to $0.05~{\rm dex}$ of the scatter across the considered halo mass range. More so in H-AGN satellites, $R_{\rm h}$ and $c_{\rm h}$ reduce scatter across all halo masses, particularly at lower halo masses (up to $0.2~{\rm dex}$), with similar behaviour observed in TNG. Therefore, even though the bivariate scatter $\sigma_{\log M_\star | \log M_{\rm h}}$ in satellites is larger than in centrals, this can be attributed to residual correlations with secondary halo properties such that the scatter after taking these into account is similar.

\subsubsection{Predicting the gas mass} \label{sec:mgas_pred}

We now move on to predicting the total gas mass, $M_{\rm gas}$. Using all informative halo features we find $R^2$ of $0.86$/$0.89$ ($0.79$/$0.70$) in H-AGN/TNG centrals (satellites). In the central panel of~\cref{fig:halo2gal_R2} we show the $R^2$ score as a function of the cumulative feature set that maximises the $R^2$ increments. As with $M_\star$ (and throughout), the most important halo feature is $M_{\rm h}$.

In H-AGN central galaxies the two important features for predicting $M_{\rm gas}$ beyond halo mass are $c_{\rm h}$ and $\lambda_{\rm h}$, whereas in TNG centrals the only other relevant feature is $\lambda_{\rm h}$. In both cases the importances of beyond-$M_{\rm h}$ properties are however relatively low ($\Delta R^2 \lesssim 0.02$). For satellites, we find that in H-AGN the other important features are $R_{\rm h}$, $M_{\rm vir}^{\rm P}$ and $c_{\rm h}$, while in TNG these are $M_{\rm vir}^{\rm P}$, $\lambda_{\rm h}$, $R_{\rm h}$ and $c_{\rm h}$. In H-AGN, $R_{\rm h}$ is the second most important feature ($\Delta R^2 \approx 0.15$), whereas in TNG its importance is near-negligible after accounting for $M_{\rm vir}^{\rm P}$ and $\lambda_{\rm h}$. As we describe in~\Cref{sec:Mass_Radius}, in H-AGN $R_{\rm h}$ is the radius of the outermost shell within which the virial theorem remains satisfied. As a consequence, for satellite galaxies $R_{\rm h}$ is not uniquely determined by $M_{\rm h}$. On the other hand, in TNG we define $R_{\rm h}$ to be distance to the furthest gravitationally bound particle and, thus, it is not uniquely determined by $M_{\rm h}$ in either parent haloes or subhaloes. This makes it potentially informative even for a model that has already been trained on $M_{\rm h}$.

\begin{figure*}
    \centering
    \includegraphics[width=\textwidth]{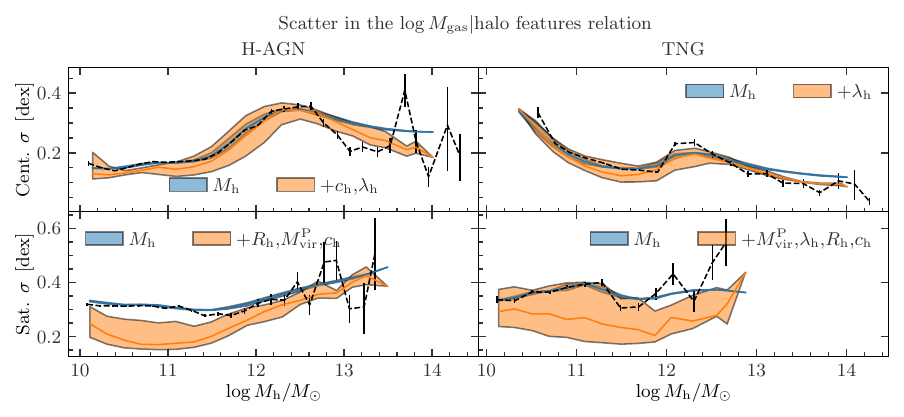}
    \caption{\acp{NN}' predicted scatter in $\log M_\mathrm{gas}$ binned by $\log M_\mathrm{h}$ in H-AGN (\emph{left} column) and TNG (\emph{right} column) for central (\emph{top} row) and satellite galaxies (\emph{bottom} row). The figure uses the same notation as~\cref{fig:NNscatter_Mstar}. In central galaxies the addition of halo properties beyond $M_{\rm h}$ yields no substantial reduction of scatter, while in satellite galaxies the scatter is reduced only marginally.}
    \label{fig:NNscatter_Mgas}
\end{figure*}

We calculate the bivariate scatter in the $\sigma_{\log M_\mathrm{gas} | \log M_{\rm h}}$ relation and plot it in~\cref{fig:NNscatter_Mgas} as the black dashed line. Focusing on H-AGN central galaxies first, we find a peak in scatter of $\sim 0.35~\mathrm{dex}$ at $M_{\rm h}\approx 10^{12.5} M_\odot$. In TNG central galaxies the scatter similarly peaks at this halo mass, though its scatter is lower by up to $0.1~{\rm dex}$. At this mass the contributions of the hot diffuse and star forming together with the cold gas are approximately equal and, since both trace $M_{\rm h}$ with a different slope and scatter, the total scatter is inflated. Below this mass the hot diffuse gas component is typically subdominant. The location of the peak is consistent with the halo mass at which AGN feedback first dominates galaxy evolution;~\citet*[see Figure 7]{Anbajagane_2022} confirm this for the TNG galaxy formation model by extracting the energetics of different baryonic processes (gas cooling, star formation, and AGN feedback) as a function of halo mass. The presence of AGN feedback is known to stop gas cooling within the halo and also expel some of the gas from the halo potential well. In TNG centrals we also note a large increase in scatter at low halo masses due to outlier simulated galaxies with artificially low star forming gas, likely due to the simulation's finite resolution. There is no prominent peak in either H-AGN nor TNG satellites where the scatter is approximately constant at $\sim0.35~{\rm dex}$ in both cases.

The coloured bands in~\cref{fig:NNscatter_Mgas} show the NN ensemble scatter in $\log M_{\rm gas}$ binned by halo mass. In both H-AGN and TNG we first train the NN ensemble on $M_\mathrm{h}$ only, which correctly recovers the bivariate scatter $\sigma_{\log M_\mathrm{gas} | \log M_\mathrm{h}}$. In centrals, adding the other informative properties yields no significant decrease in either H-AGN nor TNG, showing that the scatter in $\sigma_{\log M_\mathrm{gas} | \log M_\mathrm{h}}$ is not significantly attributable to residual correlations with other halo features. On the other hand, in both H-AGN and TNG satellites the addition of secondary properties can reduce the scatter by up to $0.1~{\rm dex}$.

\subsubsection{Predicting the neutral hydrogen mass}

\begin{figure}
    \centering
    \includegraphics[width=\columnwidth]{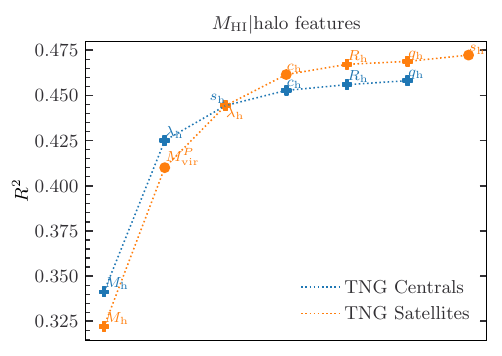}
    \caption{Test set $R^2$ scores from Extra-Trees as a function of the cumulative number of used features for predicting $M_{\rm HI}$ in TNG central and satellite galaxies. Notation follows that of~\cref{fig:halo2gal_R2}. Note that the prediction is poor ($R^2 \ll 1$) no matter how many halo features are used.}
    \label{fig:neutral_hydrogen_R2s}
\end{figure}

In TNG we also try predicting the neutral hydrogen mass $M_\mathrm{HI}$. With Extra-Trees we find maximum $R^2$ values of $0.47$ and $0.49$ in central and satellite galaxies, respectively, using all informative dark matter features. We show the $R^2$ score as a function of the cumulative feature set in~\cref{fig:neutral_hydrogen_R2s}.
In central galaxies the important halo properties besides $M_{\rm h}$ are $\lambda_{\rm h}$, $s_{\rm h}$ and $c_{\rm h}$, though it is mainly  $\lambda_{\rm h}$ that improves the predictions ($\Delta R^2 \approx 0.075$). In satellite galaxies, the relevant secondary features are $M_{\rm vir}^{\rm P}$, $\lambda_{\rm h}$ and $c_{\rm h}$. While for central galaxies $\lambda_{\rm h}$ is the second most important feature and is positively correlated with $M_{\rm HI}$ after accounting for the $M_{\rm h}$ dependence, in satellites it is the third most important feature and is negatively correlated with $M_{\rm HI}$ after accounting for both $M_{\rm h}$ and $M_{\rm vir}^{\rm P}$.

The scatter in the relation $\log M_\mathrm{HI} | \log M_{\rm h}$ is approximately constant at $0.40~\mathrm{dex}$ ($0.45~\mathrm{dex}$) in centrals (satellites), which is typically at least twice as large as the scatter in stellar mass. Because of the low $R^2$ and relatively low importance of features beyond $M_{\rm h}$, training the NN ensemble on such features can reduce the $M_{\rm HI} - M_{\rm h}$ scatter by at most $0.05~\mathrm{dex}$.

\subsubsection{Predicting the stellar half-mass radius}

Predicting the stellar half-mass radius $R_{1/2}$ from halo properties, using Extra-Trees we arrive at a maximum $R^2$ of $0.67$/$0.47$ ($0.60$/$0.41$) in H-AGN/TNG centrals (satellites) when using all informative halo features ($\Delta R^2 > 0.005$). The right column in~\cref{fig:halo2gal_R2} shows the $R^2$ scores as a function of the cumulative feature set that maximises the $R^2$ increments.

As we show in~\cref{fig:halo2gal_R2}, in H-AGN centrals the other important halo properties for predicting $R_{1/2}$ are $c_{\rm h}$ and $\lambda_{\rm h}$. In TNG centrals these are $\lambda_{\rm h}$, $s_{\rm h}$ and $c_{\rm h}$. However, in both simulations $R_{1/2}$ is poorly constrained and the addition of secondary features results in $\Delta R^2 \lesssim 0.05$. Looking at satellites, in H-AGN we see that the secondary important features are $R_{\mathrm h}$, $M_{\rm vir}^{\rm P}$, $c_{\rm h}$ and $\lambda_{\rm h}$, whereas in TNG these are $R_{\rm h}$, $\lambda_{\rm h}$, $s_{\rm h}$ and $M_{\rm vir}^{\rm P}$. In both cases the satellite $R_{1/2}$ is predicted worse than in central galaxies and $R_{\rm h}$ is the second most important feature, which is anti-correlated with $R_{1/2}$ after accounting for $M_{\rm h}$. These correlations are discussed further in the context of semi-empirical galaxy formation models in~\cref{sec:test_sam}.

\begin{figure*}
    \centering
    \includegraphics[width=\textwidth]{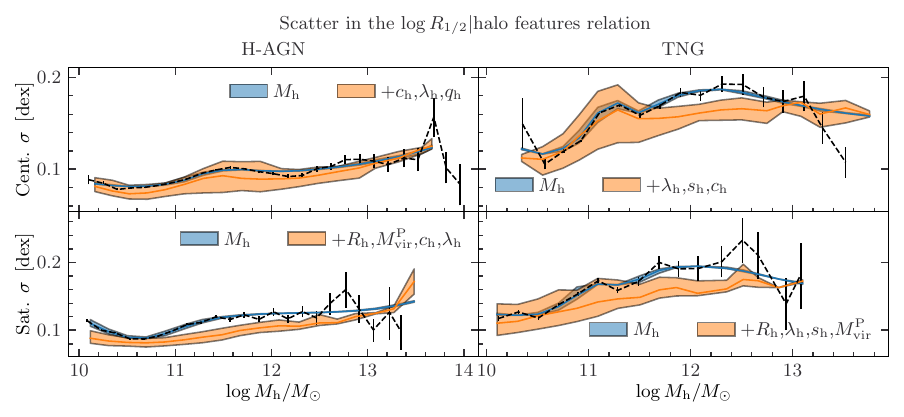}
    \caption{\acp{NN}' predicted scatter in $\log R_{1/2}$ binned by $\log M_\mathrm{h}$ in H-AGN (\emph{left column}) and TNG (\emph{right column}) for central (\emph{top row}) and satellite galaxies (\emph{bottom row}). The figure uses the same notation as~\cref{fig:NNscatter_Mstar}. Similarly as for $M_{\rm gas}$, the addition of properties beyond $M_{\rm h}$ only marginally reduces the uncertainty in the $R_{1/2}$ prediction for satellite galaxies.
    }
    \label{fig:NNscatter_Reff}
\end{figure*}

The distribution $R_{1/2} - M_\mathrm{h}$ differs in H-AGN and TNG (see~\cref{fig:sim_data}) and, as a consequence, so does the scatter $\sigma_{\log R_{1/2} | \log M_\mathrm{h}}$ which we show in~\cref{fig:NNscatter_Reff}. In H-AGN the scatter is constant at $\sim 0.1~\mathrm{dex}$ in both centrals and satellites, though the scatter in satellites is higher by $\sim 10$ per cent. On the other hand, in TNG the distribution is wider, with the scatter peaking at $\sim 0.18~\mathrm{dex}$ at $M_\mathrm{h}\approx 10^{12} M_\odot$. This peak can be attributed to the fact that above it the galaxy stellar size is proportional to the halo virial mass with a relatively small scatter, whereas below it the relation's slope flattens and widens. At the peak's location the distribution of galaxy size at fixed virial mass has a heavy tail, deviating from a Gaussian distribution. This peak is also consistent with the AGN feedback discussed above.

We train the NN ensemble on the previously identified important features and show the predicted scatter in $\log R_{1/2}$ in~\cref{fig:NNscatter_Reff}. In central galaxies calculating the NN scatter on all important halo properties yields no significant improvement over the bivariate scatter $\sigma_{\log R_{1/2} | \log M_{\rm h}}$. In H-AGN satellites the network learns to attribute $\sim10$ per cent of the scatter to secondary halo properties, however in TNG satellites the inclusion of such features does not reduce the scatter.

\subsection{Galaxy to halo mapping}
\label{sec:galaxy_to_halo}

Our machinery may easily be adapted to learn the inverse of the galaxy--halo connection as we have considered it above, i.e. the mapping from galaxy to halo properties. We consider two sets of galaxy properties as features. The first set includes all galaxy properties summarised in~\Cref{tab:properties} except absolute magnitudes, while the second consists only of absolute magnitudes as the directly observable properties. With the first set we predict both the halo mass $M_{\rm h}$ and spin $\lambda_{\rm h}$, whereas with the second we predict only $M_{\rm h}$. We briefly discuss predicting the halo concentration $c_{\rm h}$ in~\Cref{sec:Discussion}.

As in~\Cref{sec:halo_to_galaxy}, we first train an Extra-Trees algorithm to determine the important galaxy features and then the NN ensemble to learn the scatter in the relations. We compare this to the bivariate scatter in $\log M_{\rm h} | \log M_\star$. We condition this on $\log M_\star$ as we expect it \textit{a priori} to be the most important property.

\begin{figure*}
    \centering
    \includegraphics[width=\textwidth]{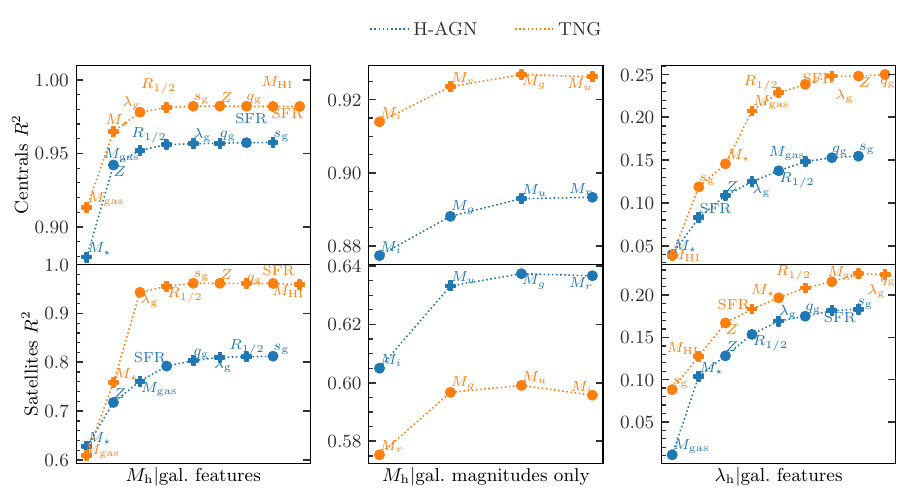}
    \caption{Test set $R^2$ scores from Extra-Trees as a function of the cumulative number of features used to predict $M_{\rm h}$ from all simulation features and from magnitudes only (\emph{left} and \emph{central} columns, respectively), and to predict $\lambda_{\rm h}$ (\emph{right} column), for central (\emph{top} row) and satellite (\emph{bottom} row) galaxies. The notation follows~\cref{fig:halo2gal_R2}.
    The magnitudes-only predictions of $M_{\rm h}$ are substantially worse in satellite galaxies and in all cases we find $\lambda_{\rm h}$ to be very poorly predicted ($R^2 \lesssim 0.25$).}
    \label{fig:gal2halo_R2s}
\end{figure*}

\subsubsection{Predicting the halo mass}\label{sec:predMhalo}

Training Extra-Trees on the galaxy properties to predict the halo mass $M_{\rm h}$ yields a maximum $R^2$ of $0.95$/$0.98$ ($0.80$/$0.95$) in H-AGN/TNG centrals (satellites) when all informative features are considered. In~\cref{fig:gal2halo_R2s} we show the $R^2$ scores as a function of the cumulative feature set.

As we show in~\cref{fig:gal2halo_R2s}, in H-AGN centrals the important galaxy properties beyond $M_\star$ are $Z$ and $M_{\rm gas}$. On the other hand, in TNG the most important property that maximises $R^2$ on its own is $M_{\rm gas}$ and the other relevant features are $M_\star$ and $\lambda_{\rm g}$. Going to satellites, in H-AGN  the most important galaxy feature is $M_\star$ and the other informative properties, ordered by decreasing importance, are $Z$, $M_{\rm gas}$, SFR, $q_{\rm g}$ and $\lambda_{\rm g}$. In TNG satellites, the most important feature is again $M_{\rm gas}$, with the other relevant features being $M_\star$, $\lambda_{\rm g}$, $R_{1/2}$ and $s_{\rm g}$. We see that in H-AGN central and satellite galaxies stellar metallicity is the second most important feature and is anti-correlated with $M_{\rm h}$ after removing the $M_\star$ dependence, and that in all cases it is beneficial to consider both $M_\star$ and $M_{\rm gas}$. This is in contrast to the halo-to-galaxy prediction where a single feature, $M_\text{h}$, was unambiguously the most important.

\begin{figure*}
    \centering
    \includegraphics[width=\textwidth]{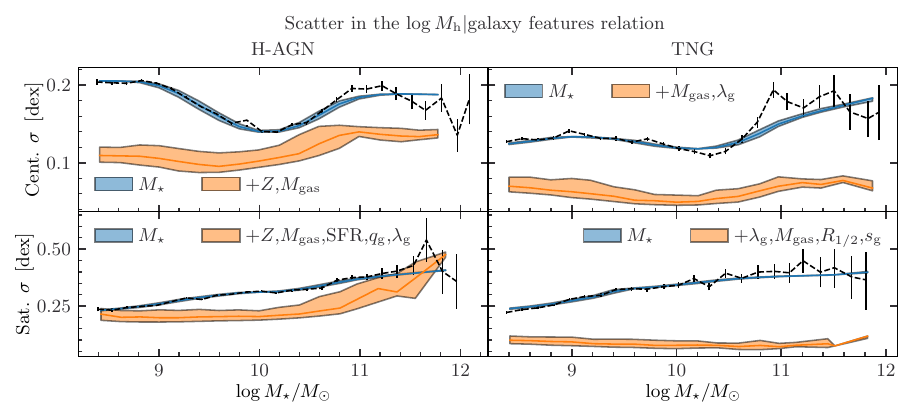}
    \caption{\acp{NN}' predicted scatter in $\log M_\mathrm{h}$ binned by $\log M_\star$ in H-AGN (\emph{left} column) and TNG (\emph{right} column) for central (\emph{top} row) and satellite (\emph{bottom} row) galaxies, training on galaxy properties. The figure's notation follows that of~\cref{fig:NNscatter_Mstar}. In all cases galaxy properties other than $M_\star$ are substantially informative for predicting $M_\text{h}$.
    }
    \label{fig:NNscatter_Mvir}
\end{figure*}

We show the bivariate scatter $\sigma_{\log M_{\rm h} | \log M_\star}$ in~\cref{fig:NNscatter_Mvir}. In H-AGN the scatter in centrals is $\sim0.2~\mathrm{dex}$ at both low and high $M_\star$, although in between the scatter decreases and at $M_\star\approx 10^{10.3} M_\odot$ it reaches $\sim0.15~\mathrm{dex}$. In TNG the scatter in centrals behaves similarly, though it is consistently lower than in H-AGN, particularly at lower stellar masses. In satellites the scatter at fixed stellar mass grows nearly monotonically from $\sim0.2~\mathrm{dex}$ at low $M_\star$ to $\sim0.4~\mathrm{dex}$ at high $M_\star$ in both H-AGN and TNG. The increase in scatter from low to high $M_\star$ can be understood from the simulated $M_\star - M_{\rm vir}$ relations (see~\cref{fig:sim_data}), which flatten at high $M_\star$. Similarly, in centrals it reaches a minimum at $M_\star \approx 10^{10.3} M_\odot$ where the two parts of the broken power law meet.

The bands in~\cref{fig:NNscatter_Mvir} show the predicted NN ensemble scatter in $\log M_{\rm h}$, binned by $M_\star$. In all cases, the networks visibly learn from properties beyond $M_\star$, since in TNG $M_{\rm gas}$ by itself is marginally even more important than $M_{\star}$. In both H-AGN and TNG central galaxies the addition of the informative features beyond $M_\star$ nearly halves the predicted scatter. In satellites the addition of the important secondary features flattens the increasing scatter at the bright end, partially in H-AGN and completely in TNG.

\subsubsection{Predicting halo mass from photometric observables}

Training Extra-Trees on the \textit{u}, \textit{g}, \textit{r} and \textit{i} magnitude bands to predict $M_{\rm h}$, we find maximum $R^2$ values of $0.89$/$0.92$ ($0.64$/$0.60$) in H-AGN/TNG central (satellite) galaxies. We show the $R^2$ as a function of the cumulative features used in~\cref{fig:gal2halo_R2s}. In all cases we find the $i$ and $r$ magnitude bands to maximise the $R^2$ increment, although the magnitudes are strongly mutually correlated. The $i$-band is the most important because it exhibits the lowest scatter in its mass-to-light ratio: in both simulations we observe that $M_\star$ correlates near-perfectly with $M_i$ (up to $\left|\rho_{\rm s}\right| \approx 0.99$) and weakest with the $u$-band magnitudes ($\left|\rho_{\rm s}\right| \lesssim 0.90$).

From the right column of~\cref{fig:gal2halo_R2s} we see again that in H-AGN centrals the best band for predicting halo mass is \textit{i}, with some additional information from the \textit{g} and \textit{u}-bands. The \textit{r}-band is considered unimportant because of its strong correlation with particularly the \textit{i}-band. In TNG the most important band is similarly \textit{i}, with only minor contribution from \textit{r}. Moving to H-AGN satellites, we see that the most important magnitude is still $M_i$ with some minor importance again in $M_u$ and  and $M_g$. In TNG satellites the most important band is now $r$, with additional minor information in \textit{g} and \textit{u}. In all cases we find that the $i$ or $r$-band magnitudes are the most important. In both simulations the direct observables capture nearly all the $M_\text{h}$ information of central galaxies, while in satellites the $R^2$ is significantly lower.

In neither case does training the \ac{NN} ensemble on more than the most important magnitude band show significant benefit in reducing the scatter in halo mass. For central galaxies the dependence of scatter on the magnitude is similar to that on stellar mass, which we show in~\cref{fig:NNscatter_Mvir}. On the other hand, in satellite galaxies the directly observable properties capture substantially less information than the full set ($\Delta R^2 \approx 0.3$), which is in part reflected by the fact that the scatter grows near linearly towards brighter galaxies even if all magnitudes are included in the training set. The main halo mass can be predicted from magnitudes with an uncertainty of $\sim0.15~\mathrm{dex}$, whereas the subhalo mass scatter runs from $\sim0.2~\mathrm{dex}$ at the faint end to $\sim0.4~\mathrm{dex}$ at the bright end.

\subsubsection{Predicting the halo spin}

Finally, we ask whether the halo spin can be inferred from galactic properties only. We find that if we consider all informative galaxy properties the maximum $R^2$ is $0.15$/$0.25$ ($0.18$/$0.23$) in H-AGN/TNG central (satellite) galaxies, suggesting that in all cases $\lambda_{\rm h}$ is very poorly predicted. We show the $R^2$ as a function of the cumulative feature set in the right panel of~\cref{fig:gal2halo_R2s}.

Starting with central galaxies, the most important properties are $M_\star$ and $M_{\rm HI}$ in H-AGN and TNG, respectively, although their $R^2$ are close to $0$, meaning that having access to only them the model at best learns to predict the average spin. While in H-AGN centrals no additional property yields a significant $R^2$ increase on its own, in TNG the addition of $s_{\rm g}$ after $M_{\rm HI}$ increases the $R^2$ by $\approx 0.08$. Similar behaviour is observed with satellites, although in the case of TNG we find $s_{\rm g}$ to be the best predictor on its own with $R^2\approx0.1$. Because of the low $R^2$ score even when training on all galaxy properties we do not pursue modelling the scatter in the galaxy--halo spin connection.


\section{Implications for semi-empirical models}
\label{sec:test_sam}

Semi-empirical and semi-analytic models of galaxy formation stipulate correlations between galaxy and halo properties which may or may not be present in the full-physics results of hydrodynamical simulations. Provided they accurately describe the real Universe, such simulations therefore provide an opportunity to test the models. Our ML analysis contributes to this by uncovering the most significant correlations of the simulations' galaxy--halo connections, while more detailed information can be obtained by testing the models explicitly on the simulated data. Here we focus on two classes of semi-empirical model: 1) subhalo abundance matching (SHAM) as a method for associating luminosity or stellar mass to haloes given their mass and mass accretion history, and 2) models that attempt to predict galaxy size from halo properties, in particular the angular momentum partition model of~\cite{Fall_Efstathiou_1980,MMW_1998} and the galaxy size--virial radius relation of ~\cite{Kravtsov_2013}.

\subsection{Subhalo abundance matching}\label{sec:abundance_match}

SHAM posits a nearly monotonic relation between the luminosity or stellar mass of a galaxy and a halo quantity called the ``proxy'' which is typically a mass or rotation velocity at a particular epoch of the halo's formation. While the earliest models used present-day virial mass as the proxy~\citep{Conroy2006, Moster2010, Behroozi_2010}, more recent work has shown that fits to galaxy clustering are improved by instead using a quantity that scales with haloes' concentration~\citep{Reddick2013, Lehmann2017}; this reflects, through the correlation of concentration and halo formation time, the fact that galaxies formed in haloes at $z>0$. SHAM also provides reasonable fits to galaxies' dynamical scaling relations, although there are indications that the galaxy mass--halo mass relation it implies is too curved and that the neglect of additional correlations in the galaxy--halo connection generates excess scatter in circular velocity at fixed mass~\citep{Trujillo, Desmond_Wechsler_2015, Desmond_Wechsler_2017, Desmond_MDAR, Desmond_BTFR, Li_2022}.

In~\cref{fig:halo2gal_R2} we show the halo features that galaxy stellar mass most strongly depends on in the H-AGN and TNG simulations. While the halo mass is clearly the strongest predictor, we find that for satellites $R_{\rm h}$ is a more important feature than concentration, indicating that SHAM may be able to provide an improved fit to the galaxy--halo connections in the simulations by incorporating this variable. The fact that $M_{\rm HI}$ is relatively poorly predicted even when all our halo properties are considered (\cref{fig:neutral_hydrogen_R2s}) explains why SHAM models have found little success in modelling the neutral hydrogen gas of galaxies~\citep{Guo, Stiskalek2021, Dutta2021}.
The models require a large scatter in the SHAM matching to replicate the poor correlation of neutral hydrogen galaxy mass with halo properties. Our results indicate that this situation could not be greatly improved no matter which additional halo variables considered in this work were included in the SHAM proxy, and also that the optimal model likely depends much more strongly on simulation-specific physics than does one based on $M_\star$. Of course we calculate the correlations only at $z=0$; stronger correlations may be found by considering halo properties e.g. at the epoch of peak mass~\citep{McGibbon}.

Our analysis also provides insight into ``inverse SHAM'', where galaxy properties are used to estimate the halo mass of an object. From~\cref{fig:gal2halo_R2s} we see that, as expected, $M_\star$ in H-AGN and $M_{\rm gas}$ in TNG are the best predictors of $M_{\rm h}$ for centrals and satellites. The gas mass tracks the halo mass particularly well in massive haloes where the gas-to-halo mass approaches the ratio of the cosmic baryon-to-dark matter density. Interestingly, we find stellar metallicity $Z$ to be the next most important variable for both satellites and centrals in H-AGN, while stellar mass plays this role in TNG. Especially for satellite galaxies these secondary correlations can be relatively important, indicating significant galaxy assembly bias. Our ML model for performing the function of SHAM or inverse SHAM would outperform these models by a considerable margin for satellites, suggesting the need for refinement of the model to reproduce the results of hydrodynamical simulations with high fidelity. This could be achieved within the framework of ``conditional abundance matching''~\citep{CAM}.

\subsection{Galaxy size models}\label{sec:galaxy_size_models}

Next we consider models for galaxy size, which traditionally employ assumptions about angular momentum (AM). Specifically, the assumption that galaxy and halo specific angular momenta are comparable has been a staple of semi-analytic modelling for over two decades, and would be expected in the absence of significant transfer or loss of AM during galaxy formation. The model may be generalised by taking the specific AM of galaxies and haloes to be proportional rather than identical, where the coefficient of proportionality $j_{\rm d}$ would describe a quasi-universal transfer of AM on average. This allows galaxy disk sizes $R_{1/2}$ to be calculated as a function of halo spin $\lambda_{\rm h}$, halo mass $M_{\rm h}$ and flat rotation curve velocity $V_{\rm c}$:
\begin{equation}\label{eq:MMW}
    R_{1/2} = \frac{\lambda_{\rm h} G M_{\rm h}^{3/2}}{2 V_{\rm c} |E|^{1/2}}\left(\frac{j_{\rm d}}{m_{\rm d}}\right) \approx \frac{j_d}{m_d} \lambda_{\rm h} R_{\rm h},
\end{equation}
where $m_{\rm d}$ is a fixed fraction of the virial mass that settles into the galaxy disk and $E$ is the total energy of the halo~\citep{MMW_1998}. Using a realistic $M_\star-M_\text{vir}$ relation it has been shown that taking $j_{\rm d}\simeq0.6-1$ gives the correct normalisation for the $M_\star-R_{1/2}$ relation~\citep{Desmond_Wechsler_2015, Posti, Pina}, although it overpredicts the scatter in sizes at fixed stellar mass and implies an unobserved strong anticorrelation between residuals of the Tully--Fisher and mass--size relations~\citep{de_Jong_2000,Gnedin_2007,Desmond_Wechsler_2015,Desmond_2019}. The model of~\citet{Kravtsov_2013} is motivated by that of~\citeauthor{MMW_1998} although it is formally separable from it: it suggests that $R_{1/2}$ is $\sim15$ per cent of $R_\text{200}$ regardless of $M_\star$. Unlike~\citeauthor{MMW_1998}, this may but does not have to imply a correlation between $R_{1/2}$ and $\lambda_{\rm h}$.

Several studies have tested the~\citeauthor{MMW_1998} model against hydrodynamical simulations, typically finding far less correlation between $R_{1/2}$ and $\lambda_{\rm h}$ than the model predicts~\citep{Sales_2009, Sales_2012, Stevens, Desmond_2017,Jiang_2019,Yang_2021,Rohr}. This accords with the empirical deficiencies of the model as described above. Our~\cref{fig:halo2gal_R2} shows that the prediction of $R_{1/2}$ is little helped by knowledge of $\lambda_{\rm h}$: in H-AGN it is the third most important halo feature for centrals and the fifth for satellites, while for TNG, although it is the second and third most important variable respectively, the $R^2$ score of the $R_{1/2}$ prediction never reaches $0.5$. This implies that galaxy size is not strongly correlated with \emph{any} halo property, indicating no significant correlations of $R_{1/2}$ beyond the $R_{1/2}-M_\star$ relation. This was also found in the EAGLE simulation~\citep{Desmond_2017}, and argued for on the basis of galaxies' dynamical scaling relations in~\citet{Desmond_MDAR} and~\citet{Desmond_2019}. In addition, Fig.~\ref{fig:gal2halo_R2s} shows that $\lambda_h$ cannot be predicted reliably from any set of galaxy properties.

\begin{figure}
    \centering
    \includegraphics[width=\columnwidth]{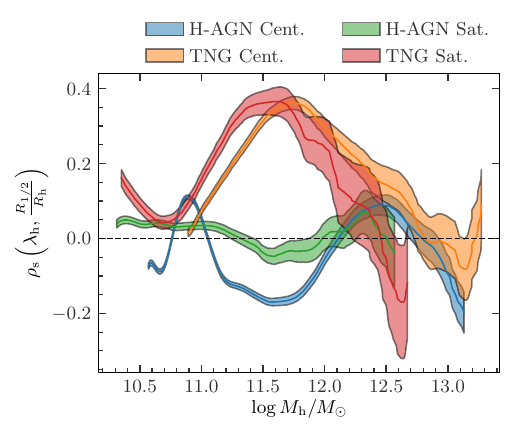}
    \caption{Spearman correlation coefficient $\rho_s$ between the halo spin $\lambda_{\rm h}$ and the galaxy-to-halo size ratio $R_{1/2}/R_{\rm h}$ as a function of $\log M_{\rm h}$ in H-AGN and TNG central and satellite galaxies. $\rho_s$ is calculated in a moving window of width $0.5~\mathrm{dex}$ in $\log M_{\rm h}$. Shaded bands indicate the $1$ $\sigma$ uncertainties in bootstrap resamples.}
    \label{fig:mvir_spearms}
\end{figure}

To test this further, in~\cref{fig:mvir_spearms} we calculate the Spearman coefficient of the $\lambda_{\rm h}-R_{1/2}/R_{\rm h}$ relation in a moving bin of $\log M_{\rm h}$ of width $0.5~\mathrm{dex}$, separately for centrals and satellites in each simulation. It is expected that $\lambda_{\rm h}$ has a linear correlation with $R_{1/2}/R_{\rm h}$ in the model of~\citeauthor{MMW_1998} (\cref{eq:MMW}), but instead we find that the correlation is always weak ($\rho_{\rm s} < 0.4$). Indeed, in H-AGN it is \emph{negative} for centrals at $M_{\rm h} \approx 10^{12} M_\odot$, indicating that faster-spinning haloes host relatively \emph{smaller} galaxies.

\begin{figure}
    \centering
    \includegraphics[width=\columnwidth]{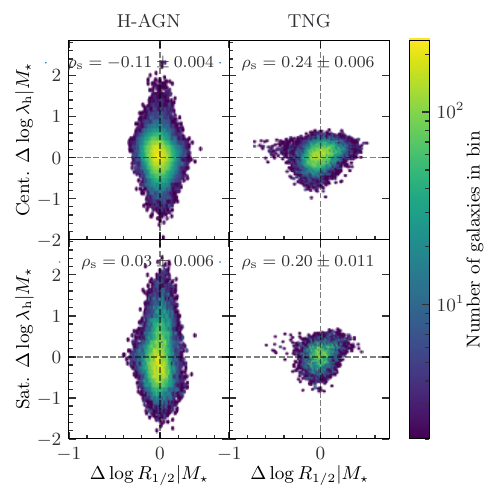}
    \caption{Correlation between the residuals of the $\lambda_{\rm h}-M_\star$ and $R_{1/2}-M_\star$ relations in H-AGN (\emph{left} column) and TNG (\emph{right} column) central (\emph{top} row) and satellite (\emph{bottom} row) galaxies. These residuals would be strongly correlated in the \citet{MMW_1998} model. Within each panel we indicate the Spearman correlation coefficient and its uncertainty estimated by bootstrapping.
    }
    \label{fig:spin_galaxysizeCorr}
\end{figure}

\begin{figure}
    \centering
    \includegraphics[width=\columnwidth]{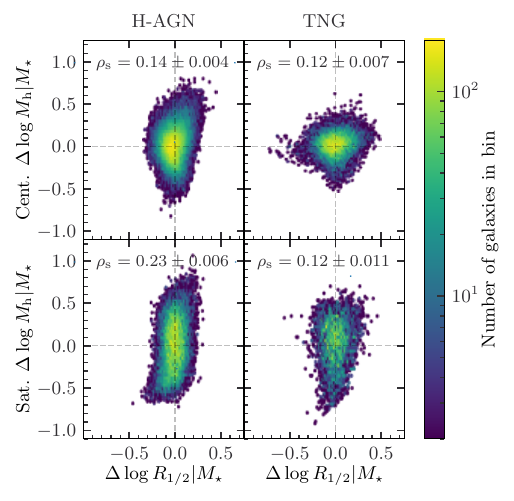}
    \caption{Same as~\cref{fig:spin_galaxysizeCorr}, except for the $M_{\rm h}-M_\star$ and $R_{1/2}-M_\star$ residual correlations. These would be strongly correlated in the~\citet{Kravtsov_2013} model.}
    \label{fig:MvirMstarReff_corr}
\end{figure}

Finally, we consider the correlations of the residuals of the $R_{1/2} - M_\star$, $\lambda_{\rm h} - M_\star$ and $M_{\rm h} - M_\star$ relations to remove the mean trend of these variables with $M_\star$. The residuals are defined by
\begin{equation}
    \Delta \log X|M_\star
    \equiv
    \log(X) - \langle \log(X)|\log(M_\star) \rangle,    
\end{equation}
where $\langle \log(X)|\log(M_\star) \rangle$ is the expectation for $\log(X)$ at given $\log(M_\star)$ from fitting a shallow forest of decision trees on subsets of the entire data set.
We plot the resulting correlations for of $\Delta \log \lambda_{\rm h}|M_\star - \Delta \log R_{1/2}|M_\star$ in~\cref{fig:spin_galaxysizeCorr} and $\Delta \log M_{\rm h}|M_\star - \Delta \log R_{1/2}|M_\star$ in~\cref{fig:MvirMstarReff_corr}, indicating in each panel the Spearman's rank correlation coefficient and its uncertainty from bootstrap resamples of the data. We find no particularly significant correlations in either simulation, in contrast to the expectations from the models of~\citeauthor{MMW_1998} and~\citeauthor{Kravtsov_2013}, respectively.


We conclude that, in accordance with observations and other hydrodynamical simulations, the model of~\citeauthor{MMW_1998} is not at play in H-AGN, while in TNG the correlation strengths fall far short of those expected from angular momentum partition. We also find that the model of~\citeauthor{Kravtsov_2013} essentially reduces to the $R_\text{eff}-M_\star$ plus $M_\star-M_\text{vir}$ relations; this is further shown by the fact that the linear correlation $R_{1/2} - M_\star$ is considerably stronger than that of $R_{1/2} - M_{\rm h}$, although their Spearman correlation coefficients are comparable due to the strong $M_\star-M_\text{h}$ relation. Lastly, we consider predicting the residuals $\Delta \log R_{1/2}|M_\star$ from the halo properties, finding that in all cases the $R^2$ scores of the predictions are $\lesssim 0.1$. A corollary is that $R_{1/2}$ provides little to no useful information for inferring $M_{\rm h}$ except through its correlation with $M_\star$, as also evidenced by~\cref{fig:halo2gal_R2} (contra~\citealt{Rohr}). More detailed numerical studies of angular momentum transport in forming galaxies are needed to provide a physical understanding of these results.

\section{Discussion}
\label{sec:Discussion}

\subsection{The galaxy--halo connections in the Horizon-AGN and Illustris-TNG simulations}\label{sec:sim_comp}

The set of galaxy and halo properties from a simulation populates a region of a high-dimensional space. As a simple example, consider the 3-dimensional space of $M_\star - M_{\rm h} - c_{\rm h}$, with $M_\star$ considered as the dependent variable. A typical, point-wise supervised ML algorithm learns the best-fit surface parameterised by $M_{\rm h}$ and $c_{\rm h}$ on which to find values of $M_\star$ consistent with the training data, such that the true values of $M_\star$ scatter minimally but by an unknown amount around it. On the other hand, our NN ensemble with a Gaussian loss function quantifies this scatter in addition to the optimal manifold, which it learns from the training data of simulated points in the $M_\star - M_{\rm h} - c_{\rm h}$ space. This allows our method to capture fully the range of possible values of $M_\star$, by means of a (Gaussian) probability distribution, from unseen $M_{\rm h}$ and $c_{\rm h}$ values.

The scatter that we are interested in learning represents the statistical uncertainty that describes the stochasticity of the simulated galaxy--halo connections, and which will tend towards a constant, typically non-zero value in the limit of infinite number of training samples. This is approximately summed in quadrature with an additional uncertainty due to the finite number of training samples. Going back to the example of the optimal manifold, parameterised by $M_{\rm h}$ and $c_{\rm h}$, the statistical uncertainty represents the scatter around the best-fit manifold.
By means of bootstrap resampling and training the models on subsets of the data we have verified that the contribution to the scatter from the finite size of the training set is subdominant to the intrinsic scatter. An important source of systematic uncertainty is the dependence of the results on the astrophysical and cosmological parameters assumed in the simulations. While we have some information on this from our use of two different simulations, a systematic investigation is lacking (see also~\cref{sec:future}).

We have trained our models on informative features only, by adding a feature only if it improves the $R^2$ score on the test set
by at least $\Delta R^2 = 0.005$. In~\cref{fig:halo2gal_R2,fig:gal2halo_R2s,fig:neutral_hydrogen_R2s} we observe that training a model on additional, non-informative features may marginally worsen the goodness-of-fit, despite Extra-Trees being known for resistance to the presence of non-informative features. Furthermore, with our NN ensemble we observe that the variation between  the models artificially increases, with the NNs being less likely to converge to a ``good'' minimum of the loss function if trained on uninformative features as well. This makes it preferable to invest time in determining which features are important for predicting a given target to achieve optimal performance, as opposed to inputting all available features into the model, many of which may act only as noise with respect to the target quantity. We now discuss our results for learning to predict $M_\star$, $M_{\rm gas}$, $R_{1/2}$, $M_{\rm HI}$ and $M_{\rm h}$.

In~\cref{fig:halo2gal_R2} we show that $M_\star$ can be predicted from the halo properties in both central and satellite galaxies with similar, high accuracy ($R^2 > 0.9$), which is consistent with previous studies~\citep{Kamdar_2016,de_Santi_2022,Lovell_2021,Chen_2021}. While in H-AGN the only relevant property beyond $M_\text{h}$ is $c_{\rm h}$, yielding $\Delta R^2 \approx 0.07$, in TNG the combination of $\lambda_{\rm h}$, $R_{\rm h}$ and $c_{\rm h}$ is required. On the other hand, in satellite galaxies we find that in both simulations the halo size $R_{\rm h}$ along with $c_{\rm h}$ is required to match the goodness-of-fit of central galaxies. Secondary halo features account for $\sim50$ per cent of the scatter in the $M_\star - M_{\rm h}$ relation as shown in~\cref{fig:NNscatter_Mstar}, indicating that their inclusion may benefit semi-empirical models such as SHAM, especially in satellites. That the features required to predicted masses are similar for both simulations is consistent with other work~\citep{Villaescusa_Navarro_2021,Chen_2021}, although we find that the relevant features for other properties (e.g. $R_{1/2}$) are simulation-dependent.

In both H-AGN and TNG central galaxies the total gas mass $M_{\rm gas}$ is well constrained from $M_{\rm h}$ alone ($R^2\approx0.87$), with the importance of other features being near-negligible ($\Delta R^2\approx0.02$). In satellite galaxies secondary properties are important, allowing $\Delta R^2 \approx0.2$. We show the importance of such features in~\cref{fig:halo2gal_R2}, although these can only explain some of the $M_{\rm h}-M_{\rm gas}$ scatter in satellite galaxies (\cref{fig:NNscatter_Mgas}). In all cases we find the stellar mass to be better constrained than the total gas mass of galaxies from the $z=0$ halo properties.

Predictions of the stellar half-mass radius $R_{1/2}$ are strongly simulation-dependent as the halo-to-$R_{1/2}$ mapping in H-AGN yields $R^2$ values higher by up to $0.2$ than in TNG, and the simulations have different secondary feature importances. This indicates a lack of convergence of state-of-the-art hydrodynamical simulations in the physics relevant to galaxy sizes. In both simulations $R_{1/2}$ is however poorly constrained, as was also found in the IllustrisTNG300 simulation~\citep{de_Santi_2022}. We find that no property beyond $M_{\rm h}$ can account for a significant amount of the $R_{1/2} - M_{\rm h}$ scatter and that $R_{1/2}$ can be predicted primarily because of its correlation with $M_{\star}$, which in turn correlates with $M_{\rm h}$. We discuss this in further detail in~\Cref{sec:test_sam}. We also study the neutral hydrogen mass relation in TNG, finding it to be poorly constrained from ($z=0$) halo features alone ($R^2 \leq 0.5$), though halo spin $\lambda_{\rm h}$ and, in the case of satellites, the parent halo virial mass are significant (\cref{fig:neutral_hydrogen_R2s}). It is possible that the halo-to-$R_{1/2}$ or $M_{\rm HI}$ connections could be better constrained if the feature set was augmented with environmental properties and/or halo variables at $z>0$.

As expected from the results above, we find the central-galaxy-to-$M_{\rm h}$ connection to be near perfectly constrained in both H-AGN and TNG ($R^2 \approx 0.95, 0.98$, respectively). The satellite connection is not as well constrained ($R^2 \approx 0.8, 0.9$), however, as we show in~\cref{fig:gal2halo_R2s}. Properties beyond $M_\star$ account for up to $50$ per cent of the scatter in the $M_{\rm h} - M_\star$ relation (\cref{fig:NNscatter_Mvir}), such that the central and satellite halo masses can be constrained with an uncertainty of $\sim~0.1~\mathrm{dex}$ and $\sim~0.1-0.2~\mathrm{dex}$, respectively. This is comparable to the scatter found in similar studies which analyse the CAMELS simulations~\citep{Villanueva_Domingo_2021,Shao_2021}. The central halo mass can be nearly as well predicted from magnitudes only, although in satellites the predictions are again worse (\cref{fig:gal2halo_R2s}). We find the $i$-band magnitude to be the most important, as expected from its strongest correlation with $M_\star$ compared to the other bands. Although, if magnitudes only are used as inputs, we find that the scatter in $M_{\rm h}$ starts significantly growing towards brighter galaxies. The predictive power of raw photometry is comparable to that found by~\cite{vonMarttens_2021}, who also studied the TNG simulations but with a Tree-based Pipeline Optimization Tool~\citep{Olson_2016}. They found that the total mass could be predicted from magnitudes with $R^2 \lesssim0.86$ and from all features with $R^2 \approx 0.98$, again in agreement with our results (see~\Cref{sec:predMhalo}).

We also consider predicting the halo spin, finding in both simulations that no combination of galaxy properties can predict $\lambda_{\rm h}$ to better than $R^2 \approx 0.25$. \cref{fig:halo2gal_R2} shows that the stellar half-mass radius $R_{1/2}$ is never a particularly important feature for predicting $\lambda_{\rm h}$. This provides further evidence that the model of~\citealt{MMW_1998} is not present in either H-AGN or TNG, which implies that $R_{1/2}$ should be proportional to $\lambda_{\rm h}$ (\cref{eq:MMW}).

Lastly, we also test learning to predict the halo concentration $c_{\rm h}$ from galaxy properties, finding that the results are strongly dependant on the concentration definition. In H-AGN the concentration is poorly constrained from the galaxy features with $R^2 \approx 0.53, 0.38$ in central and satellite galaxies, respectively. In TNG these numbers are $R^2\approx0.12, 0.32$. This suggests that in case of TNG centrals the model can do little better than predict the average value of concentration. As per~\Cref{sec:halo_concentration}, we define the concentration to be the ratio between $R_{\rm h}$ and $r_{\rm s}$. However, while in H-AGN $R_{\rm h}$ is the virial radius of the halo, in TNG it is the distance to the furthest gravitationally bound \ac{DM} particle. A consequence is that the halo concentration correlates much more weakly with halo mass in TNG than in H-AGN and is therefore less predictable from galaxy variables.

\subsection{Future work}
\label{sec:future}

\subsubsection{Applications}

Our method has a range of applications. The most obvious is to elucidate the galaxy--halo connections in hydrodynamical simulations by quantifying the intrinsic scatter of various multi-parameter correlations. This in turns promotes understanding of the effects of the feedback and subgrid prescriptions implemented in the simulations, and informs the construction of semi-empirical and semi-analytic models that attempt to capture the salient parts of this physics. In common with other machine learning analyses, our techniques also allow hydrodynamical simulation to be bypassed in connecting galaxy and halo variables. This can be used either to paste galaxies onto haloes produced in $N$-body simulations in a manner that mimics their association in full-physics simulations, or, in the opposite direction, to infer haloes from observed galaxies. The former is useful for making mock universes from cheap \ac{DM}-only simulations---which can then be used either for forecasting upcoming cosmological galaxy surveys~\citep[e.g.][]{Reddick_2014, Zhang_2019, Kasmanoff_2020, Wechsler2021Addgals} or for validating and developing analysis pipelines for these surveys \citep[e.g.][]{DeRose2019Buzzard}---while the latter is useful for estimating the distribution of \ac{DM} associated with the observed distribution of light. This has been explored in the context of predicting halo masses of an observed galaxy and group catalogue from the Sloan Digital Sky Survey, finding that it outperforms conventional halo abundance matching or dynamical mass modelling~\citep{Calderon_2019}

The advantage of using machine learning applied to simulations is that it provides a powerful way of capturing \emph{all} relevant correlations, regardless of their (unknown) functional form, at the cost of reduced generality in that the model is specific to the simulation it was trained on. We have seen important similarities and differences between the galaxy--halo connections of Horizon-AGN and IllustrisTNG, indicating that some parts of the model are robust to simulation-specific physics while others are not. In the future, techniques such as Domain Adaptation and Transfer Learning could be used to generalise the ML models and extend their functionality across multiple data sets and hence create less simulation-specific results \citep[e.g.][]{Ciprijanovic2021DomAdap}.

\subsubsection{Simulations and Features}

As described in~\cref{sec:sim_comp}, our \acp{NN} learn the high-dimensional manifold characterising the mean galactic and halo properties, and how far from this manifold objects typically reside. Since the location and shape of the manifold depends on cosmological and astrophysical parameters (see e.g.~\cite{Villaescusa2022SingleGalaxy}), in this work we have compared two simulations with different cosmologies and subgrid physics models.  One way to extend our framework is to consider a wider range of training sets which systematically vary such differences, which would improve understanding of the sensitivity of the results to the simulation inputs. A useful meta-sample comes from the CAMELS project~\citep{CAMELS_2021,CAMELS_2022}, a suite of $2,000$ magneto-hydrodynamic simulations with varying initial conditions, cosmological and baryonic parameters, and two different subgrid physics models, designed  with machine learning applications in mind (e.g.~\citealt{Villaescusa_Navarro_2021_marg}). The CAMELS boxes are however small ($25~\mathrm{Mpc}/h$ in side length), increasing sample variance.

Other existing simulation suites, of large box sizes with adequate resolution for studies of more massive haloes, already contain multiple runs that vary input parameters for AGN feedback~\citep[e.g.][]{McCarthy2017Bahamas} and cosmology (e.g. Magneticum Pathfinder\footnote{\url{http://www.magneticum.org/}}), or vary whole galaxy formation models~\citep[eg.][]{Cui2018The300}. Ensembles constructed by combining different suites have already been used to study the impact of galaxy formation (or lack thereof) on some galaxy/halo observables~\citep[e.g.][]{Anbajagane2020Stellar, Lim2021Gas, Ragagnin2021HOD, Anbajagane2022GalaxyVel}, and such ensembles also have the potential to serve as training sets. An obvious disadvantage here however is that while CAMELS provides a homogeneous parametrisation for feedback, an ensemble constructed from disparate simulations does not. In principle, an ``effective'', common parametrisation of the simulation suites is possible, but given basic uncertainties in small-scale baryonic physics and its implementation this would not be simple to construct.

Future work should consider whether the inclusion of other galaxy or halo variables, such as summaries of the merger history as in \cite{Yongseok_2019}, could further improve the NN's predictions. Also interesting is how the scatter in the galaxy--halo connection evolves as a function of redshift. Studies of the Illustris-TNG simulation using Extremely Randomised Trees~\citep{McGibbon} have found that it is possible to make significantly more accurate predictions by including features from a range of redshifts. In particular, low-redshift observables are most important for determining the gas mass and star formation rate, although predictions of stellar mass and metallicity benefit from higher redshift information. There are hints of universal relations between total halo mass and galaxy properties which are independent of redshift~\citep{Shao_2021}; it would be interesting to see whether such relations exist for the scatter as well as the mean prediction.

Finally, we have not considered environmental properties of the galaxy or halo in this work, and have thus treated all objects equally, independent of their location in the simulation box. Since it has been shown that environmental overdensity and shear can improve clustering when predicting the halo occupation distribution~\citep{Delgado_2021} and that the local environment is an important feature in predicting galactic properties \citep{Agarwal_2018}, there is potentially useful information in such features which should be exploited in future work. However, we note that~\cite{Lovell_2021} report negligible improvements when predicting the majority of galaxy properties based on their host haloes in the EAGLE/C-EAGLE simulations with the local density included in the feature set.

\subsubsection{Methodology}

Although we have been able to determine which features are important for predicting certain properties and have attempted to decipher the correlations being utilised, as with many machine learning analyses interpreting the results can be challenging. Techniques such as symbolic regression, which learns analytic forms for the relationships between variables, have already been used to find analytic approximations to the mean relations between galaxy and halo properties~\citep{Shao_2021,Delgado_2021}, and may profitably be applied also to the scatter.

In~\cref{sec:sim_scatter} we outlined how we calculate the bivariate scatter in a relation $Y-X$ at fixed $X$ by fitting a univariate Gaussian distribution to samples binned by $X$. In doing so we have assumed that a Gaussian distribution is sufficient to describe the distribution of samples within each bin. However, this assumption will be violated in case of e.g. heavy tails or the presence of outliers. These would increase the best-fit scatter and make its interpretation non-trivial. Therefore, an optimal approach would allow for a more general distribution than a Gaussian. Similarly, in~\cref{eq:Gaussian loss function}, we have assumed a Gaussian loss function, although in principle the Gaussian distribution can be replaced with any probability distribution. Moment Networks~\citep{Jeffrey_2020} could be used to determine the mean and scatter (i.e. the first two moments) of these relationships in a likelihood-free manner, without assuming a Gaussian distribution.

For each base network, $q$, we find the mean, $\mu_q$, and scatter, $\sigma_q$, that maximise the likelihood, such that their combination through~\cref{eq:Combine NN} should provide a good estimate for the true maximum likelihood. However, the distributions of $\mu_q$ and $\sigma_q$ cannot be interpreted as posterior distributions for $\mu$ and $\sigma$, and thus we cannot assign an object-by-object uncertainty to the parameters $\mu$ and $\sigma$. In plots such as Fig.~\ref{fig:NNscatter_Mstar} we therefore assign uncertainties using the spread among galaxies in a given mass bin. Future work could use methods such as Bayesian Neural Networks (e.g.~\citet{BayesNN}) to determine the uncertainties on the probability distributions themselves.

We have either manually tested which hyperparameters of the NN architecture (depth, width, activation function, optimiser, etc.) represent the data the best or performed a brute-force hyperparameter grid search in case of Extra-Trees. Although we do not expect our results to be heavily architecture-dependent (the several methods we tested gave similar results), we cannot rule out the possibility that there exists some superior hyperparameter combination. In principle one should search the entire space of hyperparameters for a way to do this automatically and efficiently (see e.g.~\citealt{Optuna} and references therein).

Finally, it would be possible to predict several target variables simultaneously with a multivariate loss function, allowing the model to learn the correlations between the target variables. In combination with the above this would ultimately allow the most faithful characterisation of the galaxy--halo connection.

\section{Conclusion}\label{sec:Conclusion}

We build an ensemble of \aclp{NN} with a Gaussian loss function to study the $z=0$ galaxy--halo connection in the Horizon-AGN (H-AGN) and Illustris-TNG100 (TNG) simulations, with particular emphasis on modelling the intrinsic scatter of this connection. This is equivalent to an uncertainty in predicting a galaxy/halo property when some subset of halo/galaxy properties is known. We quantify the importance of various halo properties for the prediction of $M_\star$, $M_{\rm gas}$, $R_{1/2}$ and $M_{\rm HI}$, and of various galaxy properties in the prediction of $M_{\rm h}$ and $\lambda_{\rm h}$. We then examine how the addition of secondary properties decreases the predicted uncertainty relative to the uncertainty of a model trained on the most important property only. This enables us to quantify, for example, how much scatter in the $M_\star - M_{\rm h}$ relation can be attributed to residual correlations with e.g. the halo concentration $c_{\rm h}$, which we show in~\cref{fig:NNscatter_Mstar}. We validate our bespoke \acl{NN} against an Extra-Trees decision tree ensemble. Lastly, we explore the implications of our findings for semi-empirical galaxy formation models. We find that while subhalo abundance matching captures the most important correlations of the simulations it may be possible to improve it, and that galaxy size--halo spin (or halo size) models do not provide a good fit to the simulated data or aid in interpreting it.

Our main findings are:
\begin{itemize}
    \item While the halo-to-stellar mass relation is well-constrained in central galaxies, with a relatively low importance of any other halo quantity, properties such as the halo size and concentration play an important role in satellite galaxies. After accounting for these properties the main halo and subhalo masses can be predicted nearly equally well ($R^2 \geq 0.9$). Secondary properties account for up to $0.1~\mathrm{dex}$ and $0.2~\mathrm{dex}$ of the scatter in the $M_\star - M_{\rm h}$ relation in central and satellite galaxies, respectively.
    \item In both H-AGN and TNG, stellar mass is more predictable from halo properties than the total gas mass with $R^2$ higher by $\sim0.05$ and $\sim0.10$ in central and satellite galaxies, respectively. The halo-to-neutral hydrogen mass relation is poorly constrained with $R^2\approx0.5$.
    \item Main halo masses can be predicted near perfectly in both simulations from galaxy properties ($R^2 \geq 0.95$), although not subhalo masses. Furthermore, while the main halo mass can be predicted as well from magnitudes only, that is not the case for subhaloes. The uncertainty in the halo mass is $\sim0.1$ and $0.2~\mathrm{dex}$ in centrals and satellites, respectively, after accounting for all galaxy properties. However, if only directly observable properties (photometric magnitudes) are considered then the uncertainty grows towards the bright end, most notably in satellites where it reaches up to $\sim0.4~\mathrm{dex}$.
    \item We find that galaxy size-based semi-analytical models are not at play in either H-AGN nor TNG, although the quantitative correlations relevant to these models are simulation-dependent. In both simulations the correlation between galaxy and halo size can essentially be reduced to the separate correlations of galaxy size with stellar mass, stellar mass with halo mass and halo mass with halo size.
\end{itemize}

By studying the Horizon-AGN and TNG simulations we showcase the ability of machine learning to quantify all relevant correlations of the galaxy--halo connection, including their intrinsic scatter. By modelling the statistical uncertainty in predictions for galaxy properties based on halo ones, or vice versa, this can be used to introduce stochasticity into the generation of mock galaxy or halo samples and hence produce more realistic and useful catalogues.

\section*{Acknowledgements}

We thank Julien Devriendt, Clotilde Laigle and Benjamin P. Moster for input and discussions. We also thank Jonathan Patterson for smoothly running the Glamdring Cluster hosted by the University of Oxford, where most of the data processing was performed.

RS is supported by the Deutscher Akademischer Austauschdienst (DAAD) Study Scholarship. DJB is supported by STFC and Oriel College, Oxford. HD was supported by St John's College, Oxford, a McWilliams Fellowship at Carnegie Mellon University and a Royal Society University Research Fellowship (grant no. 211046). DA is supported by the National Science Foundation Graduate Research Fellowship under Grant No. DGE 1746045. This project has received funding from the European Research Council (ERC) under the European Union’s Horizon 2020 research and innovation programme (grant agreement No 693024).

We would like to thank the Horizon-AGN collaboration for allowing us to use the simulation data, and particularly Stephane Rouberol for smoothly running the Horizon Cluster hosted by the Institut d'Astrophysique de Paris where most of the processing of the raw simulation data was performed. Some of the numerical work also made use of the DiRAC Data Intensive service at Leicester, operated by the University of Leicester IT Services, which forms part of the STFC DiRAC HPC Facility (\url{www.dirac.ac.uk}). The equipment was funded by BEIS capital funding via STFC capital grants ST/K000373/1 and ST/R002363/1 and STFC DiRAC Operations grant ST/R001014/1. DiRAC is part of the National e-Infrastructure.

We thank the IllustrisTNG team for publicly releasing all simulation data. The IllustrisTNG simulations were undertaken with compute time awarded by the Gauss Centre for Supercomputing (GCS) under GCS Large-Scale Projects GCS-ILLU and GCS-DWAR on the GCS share of the supercomputer Hazel Hen at the High Performance Computing Center Stuttgart (HLRS), as well as on the machines of the Max Planck Computing and Data Facility (MPCDF) in Garching, Germany.

\section*{Data availability}
We make our \ac{NN} implementation publicly available at \url{https://github.com/Richard-Sti/scatternn}. The Horizon-AGN data is available from the Horizon-AGN collaboration on request. The IllustrisTNG data is publicly available at \url{https://www.tng-project.org/data/} and the public release products are described in~\citet{Nelson_2019}. All other data will be shared on reasonable request to the authors.

\appendix

\bibliographystyle{mnras}
\bibliography{ref}

\begin{thebibliography}{}
\makeatletter
\relax
\def\mn@urlcharsother{\let\do\@makeother \do\$\do\&\do\#\do\^\do\_\do\%\do\~}
\def\mn@doi{\begingroup\mn@urlcharsother \@ifnextchar [ {\mn@doi@}
  {\mn@doi@[]}}
\def\mn@doi@[#1]#2{\def\@tempa{#1}\ifx\@tempa\@empty \href
  {http://dx.doi.org/#2} {doi:#2}\else \href {http://dx.doi.org/#2} {#1}\fi
  \endgroup}
\def\mn@eprint#1#2{\mn@eprint@#1:#2::\@nil}
\def\mn@eprint@arXiv#1{\href {http://arxiv.org/abs/#1} {{\tt arXiv:#1}}}
\def\mn@eprint@dblp#1{\href {http://dblp.uni-trier.de/rec/bibtex/#1.xml}
  {dblp:#1}}
\def\mn@eprint@#1:#2:#3:#4\@nil{\def\@tempa {#1}\def\@tempb {#2}\def\@tempc
  {#3}\ifx \@tempc \@empty \let \@tempc \@tempb \let \@tempb \@tempa \fi \ifx
  \@tempb \@empty \def\@tempb {arXiv}\fi \@ifundefined
  {mn@eprint@\@tempb}{\@tempb:\@tempc}{\expandafter \expandafter \csname
  mn@eprint@\@tempb\endcsname \expandafter{\@tempc}}}

\bibitem[\protect\citeauthoryear{Abadi et~al.,}{Abadi
  et~al.}{2015}]{tensorflow2015-whitepaper}
Abadi M.,  et~al., 2015, {TensorFlow}: Large-Scale Machine Learning on
  Heterogeneous Systems, \url {https://www.tensorflow.org/}

\bibitem[\protect\citeauthoryear{{Agarwal}, {Dav{\'e}}  \& {Bassett}}{{Agarwal}
  et~al.}{2018}]{Agarwal_2018}
{Agarwal} S.,  {Dav{\'e}} R.,   {Bassett} B.~A.,  2018, \mn@doi [\mnras]
  {10.1093/mnras/sty1169}, \href
  {https://ui.adsabs.harvard.edu/abs/2018MNRAS.478.3410A} {478, 3410}

\bibitem[\protect\citeauthoryear{{Akiba}, {Sano}, {Yanase}, {Ohta}  \&
  {Koyama}}{{Akiba} et~al.}{2019}]{Optuna}
{Akiba} T.,  {Sano} S.,  {Yanase} T.,  {Ohta} T.,   {Koyama} M.,  2019, arXiv
  e-prints, \href {https://ui.adsabs.harvard.edu/abs/2019arXiv190710902A} {p.
  arXiv:1907.10902}

\bibitem[\protect\citeauthoryear{{Anbajagane}, {Evrard}, {Farahi}, {Barnes},
  {Dolag}, {McCarthy}, {Nelson}  \& {Pillepich}}{{Anbajagane}
  et~al.}{2020}]{Anbajagane2020Stellar}
{Anbajagane} D.,  {Evrard} A.~E.,  {Farahi} A.,  {Barnes} D.~J.,  {Dolag} K.,
  {McCarthy} I.~G.,  {Nelson} D.,   {Pillepich} A.,  2020, \mn@doi [\mnras]
  {10.1093/mnras/staa1147}, \href
  {https://ui.adsabs.harvard.edu/abs/2020MNRAS.495..686A} {495, 686}

\bibitem[\protect\citeauthoryear{{Anbajagane}, {Evrard}  \&
  {Farahi}}{{Anbajagane} et~al.}{2022a}]{Anbajagane_2022}
{Anbajagane} D.,  {Evrard} A.~E.,   {Farahi} A.,  2022a, \mn@doi [\mnras]
  {10.1093/mnras/stab3177}, \href
  {https://ui.adsabs.harvard.edu/abs/2022MNRAS.509.3441A} {509, 3441}

\bibitem[\protect\citeauthoryear{{Anbajagane} et~al.,}{{Anbajagane}
  et~al.}{2022b}]{Anbajagane2022GalaxyVel}
{Anbajagane} D.,  et~al., 2022b, \mn@doi [\mnras] {10.1093/mnras/stab3587},
  \href {https://ui.adsabs.harvard.edu/abs/2022MNRAS.510.2980A} {510, 2980}

\bibitem[\protect\citeauthoryear{{Aubert}, {Pichon}  \& {Colombi}}{{Aubert}
  et~al.}{2004}]{Aubert_2004}
{Aubert} D.,  {Pichon} C.,   {Colombi} S.,  2004, \mn@doi [\mnras]
  {10.1111/j.1365-2966.2004.07883.x}, \href
  {https://ui.adsabs.harvard.edu/abs/2004MNRAS.352..376A} {352, 376}

\bibitem[\protect\citeauthoryear{{Bartlett}, {Desmond}, {Devriendt}, {Ferreira}
   \& {Slyz}}{{Bartlett} et~al.}{2021}]{Bartlett_OffsetBH_2021}
{Bartlett} D.~J.,  {Desmond} H.,  {Devriendt} J.,  {Ferreira} P.~G.,   {Slyz}
  A.,  2021, \mn@doi [\mnras] {10.1093/mnras/staa3516}, \href
  {https://ui.adsabs.harvard.edu/abs/2021MNRAS.500.4639B} {500, 4639}

\bibitem[\protect\citeauthoryear{{Baugh}}{{Baugh}}{2006}]{Baugh}
{Baugh} C.~M.,  2006, \mn@doi [Reports on Progress in Physics]
  {10.1088/0034-4885/69/12/R02}, \href
  {https://ui.adsabs.harvard.edu/abs/2006RPPh...69.3101B} {69, 3101}

\bibitem[\protect\citeauthoryear{{Behroozi}, {Conroy}  \&
  {Wechsler}}{{Behroozi} et~al.}{2010}]{Behroozi_2010}
{Behroozi} P.~S.,  {Conroy} C.,   {Wechsler} R.~H.,  2010, \mn@doi [\apj]
  {10.1088/0004-637X/717/1/379}, \href
  {https://ui.adsabs.harvard.edu/abs/2010ApJ...717..379B} {717, 379}

\bibitem[\protect\citeauthoryear{{Berlind} \& {Weinberg}}{{Berlind} \&
  {Weinberg}}{2002}]{HOD}
{Berlind} A.~A.,  {Weinberg} D.~H.,  2002, \mn@doi [\apj] {10.1086/341469},
  \href {https://ui.adsabs.harvard.edu/abs/2002ApJ...575..587B} {575, 587}

\bibitem[\protect\citeauthoryear{Breiman}{Breiman}{2001}]{Breiman_2001_RandomForest}
Breiman L.,  2001, \mn@doi [Mach. Learn.] {10.1023/A:1010933404324}, 45, 5–32

\bibitem[\protect\citeauthoryear{{Bryan} \& {Norman}}{{Bryan} \&
  {Norman}}{1998}]{BryanNorman1998}
{Bryan} G.~L.,  {Norman} M.~L.,  1998, \mn@doi [\apj] {10.1086/305262}, \href
  {https://ui.adsabs.harvard.edu/abs/1998ApJ...495...80B} {495, 80}

\bibitem[\protect\citeauthoryear{{Bullock}, {Dekel}, {Kolatt}, {Kravtsov},
  {Klypin}, {Porciani}  \& {Primack}}{{Bullock} et~al.}{2001}]{Bullock_2001}
{Bullock} J.~S.,  {Dekel} A.,  {Kolatt} T.~S.,  {Kravtsov} A.~V.,  {Klypin}
  A.~A.,  {Porciani} C.,   {Primack} J.~R.,  2001, \mn@doi [\apj]
  {10.1086/321477}, \href
  {https://ui.adsabs.harvard.edu/abs/2001ApJ...555..240B} {555, 240}

\bibitem[\protect\citeauthoryear{{Calderon} \& {Berlind}}{{Calderon} \&
  {Berlind}}{2019}]{Calderon_2019}
{Calderon} V.~F.,  {Berlind} A.~A.,  2019, \mn@doi [\mnras]
  {10.1093/mnras/stz2775}, \href
  {https://ui.adsabs.harvard.edu/abs/2019MNRAS.490.2367C} {490, 2367}

\bibitem[\protect\citeauthoryear{Capano \& Stiskalek}{Capano \&
  Stiskalek}{2021}]{epsie}
Capano C.,  Stiskalek R.,  2021, cdcapano/epsie:,
  \mn@doi{10.5281/zenodo.5717226}, \url
  {https://doi.org/10.5281/zenodo.5717226}

\bibitem[\protect\citeauthoryear{{Chabrier}}{{Chabrier}}{2003}]{Chabrier_2003}
{Chabrier} G.,  2003, \mn@doi [\pasp] {10.1086/376392}, \href
  {https://ui.adsabs.harvard.edu/abs/2003PASP..115..763C} {115, 763}

\bibitem[\protect\citeauthoryear{{Chen}, {Mo}, {Li}, {Wang}, {Wang}, {Yang},
  {Zhang}  \& {Katz}}{{Chen} et~al.}{2021}]{Chen_2021}
{Chen} Y.,  {Mo} H.~J.,  {Li} C.,  {Wang} K.,  {Wang} H.,  {Yang} X.,  {Zhang}
  Y.,   {Katz} N.,  2021, \mn@doi [\mnras] {10.1093/mnras/stab2377}, \href
  {https://ui.adsabs.harvard.edu/abs/2021MNRAS.507.2510C} {507, 2510}

\bibitem[\protect\citeauthoryear{Cheng et~al.,}{Cheng
  et~al.}{2016}]{Cheng_wide_deep}
Cheng H.-T.,  et~al., 2016, in Proceedings of the 1st Workshop on Deep Learning
  for Recommender Systems. DLRS 2016.
Association for Computing Machinery, New York, NY, USA, p. 7–10,
  \mn@doi{10.1145/2988450.2988454}, \url
  {https://doi.org/10.1145/2988450.2988454}

\bibitem[\protect\citeauthoryear{Chipman, George  \& McCulloch}{Chipman
  et~al.}{2010}]{BART}
Chipman H.~A.,  George E.~I.,   McCulloch R.~E.,  2010, \mn@doi [The Annals of
  Applied Statistics] {10.1214/09-AOAS285}, 4, 266

\bibitem[\protect\citeauthoryear{{Chisari} et~al.,}{{Chisari}
  et~al.}{2015}]{Chisari_2015}
{Chisari} N.,  et~al., 2015, \mn@doi [\mnras] {10.1093/mnras/stv2154}, \href
  {https://ui.adsabs.harvard.edu/abs/2015MNRAS.454.2736C} {454, 2736}

\bibitem[\protect\citeauthoryear{{Chisari} et~al.,}{{Chisari}
  et~al.}{2017}]{Chisari_2017}
{Chisari} N.~E.,  et~al., 2017, \mn@doi [\mnras] {10.1093/mnras/stx1998}, \href
  {https://ui.adsabs.harvard.edu/abs/2017MNRAS.472.1163C} {472, 1163}

\bibitem[\protect\citeauthoryear{{{\'C}iprijanovi{\'c}}
  et~al.,}{{{\'C}iprijanovi{\'c}} et~al.}{2021}]{Ciprijanovic2021DomAdap}
{{\'C}iprijanovi{\'c}} A.,  et~al., 2021, \mn@doi [\mnras]
  {10.1093/mnras/stab1677}, \href
  {https://ui.adsabs.harvard.edu/abs/2021MNRAS.506..677C} {506, 677}

\bibitem[\protect\citeauthoryear{{Conroy}, {Wechsler}  \& {Kravtsov}}{{Conroy}
  et~al.}{2006}]{Conroy2006}
{Conroy} C.,  {Wechsler} R.~H.,   {Kravtsov} A.~V.,  2006, \mn@doi [\apj]
  {10.1086/503602}, \href
  {https://ui.adsabs.harvard.edu/abs/2006ApJ...647..201C} {647, 201}

\bibitem[\protect\citeauthoryear{{Cui} et~al.,}{{Cui}
  et~al.}{2018}]{Cui2018The300}
{Cui} W.,  et~al., 2018, \mn@doi [\mnras] {10.1093/mnras/sty2111}, \href
  {https://ui.adsabs.harvard.edu/abs/2018MNRAS.480.2898C} {480, 2898}

\bibitem[\protect\citeauthoryear{{DeRose} et~al.,}{{DeRose}
  et~al.}{2019}]{DeRose2019Buzzard}
{DeRose} J.,  et~al., 2019, arXiv e-prints, \href
  {https://ui.adsabs.harvard.edu/abs/2019arXiv190102401D} {p. arXiv:1901.02401}

\bibitem[\protect\citeauthoryear{{Delgado}, {Wadekar}, {Hadzhiyska}, {Bose},
  {Hernquist}  \& {Ho}}{{Delgado} et~al.}{2021}]{Delgado_2021}
{Delgado} A.~M.,  {Wadekar} D.,  {Hadzhiyska} B.,  {Bose} S.,  {Hernquist} L.,
   {Ho} S.,  2021, arXiv e-prints, \href
  {https://ui.adsabs.harvard.edu/abs/2021arXiv211102422D} {p. arXiv:2111.02422}

\bibitem[\protect\citeauthoryear{{Desmond}}{{Desmond}}{2017a}]{Desmond_MDAR}
{Desmond} H.,  2017a, \mn@doi [\mnras] {10.1093/mnras/stw2571}, \href
  {https://ui.adsabs.harvard.edu/abs/2017MNRAS.464.4160D} {464, 4160}

\bibitem[\protect\citeauthoryear{{Desmond}}{{Desmond}}{2017b}]{Desmond_BTFR}
{Desmond} H.,  2017b, \mn@doi [\mnras] {10.1093/mnrasl/slx134}, \href
  {https://ui.adsabs.harvard.edu/abs/2017MNRAS.472L..35D} {472, L35}

\bibitem[\protect\citeauthoryear{{Desmond} \& {Wechsler}}{{Desmond} \&
  {Wechsler}}{2015}]{Desmond_Wechsler_2015}
{Desmond} H.,  {Wechsler} R.~H.,  2015, \mn@doi [\mnras]
  {10.1093/mnras/stv1978}, \href
  {https://ui.adsabs.harvard.edu/abs/2015MNRAS.454..322D} {454, 322}

\bibitem[\protect\citeauthoryear{{Desmond} \& {Wechsler}}{{Desmond} \&
  {Wechsler}}{2017}]{Desmond_Wechsler_2017}
{Desmond} H.,  {Wechsler} R.~H.,  2017, \mn@doi [\mnras]
  {10.1093/mnras/stw2804}, \href
  {https://ui.adsabs.harvard.edu/abs/2017MNRAS.465..820D} {465, 820}

\bibitem[\protect\citeauthoryear{{Desmond}, {Mao}, {Wechsler}, {Crain}  \&
  {Schaye}}{{Desmond} et~al.}{2017}]{Desmond_2017}
{Desmond} H.,  {Mao} Y.-Y.,  {Wechsler} R.~H.,  {Crain} R.~A.,   {Schaye} J.,
  2017, \mn@doi [\mnras] {10.1093/mnrasl/slx093}, \href
  {https://ui.adsabs.harvard.edu/abs/2017MNRAS.471L..11D} {471, L11}

\bibitem[\protect\citeauthoryear{{Desmond}, {Katz}, {Lelli}  \&
  {McGaugh}}{{Desmond} et~al.}{2019}]{Desmond_2019}
{Desmond} H.,  {Katz} H.,  {Lelli} F.,   {McGaugh} S.,  2019, \mn@doi [\mnras]
  {10.1093/mnras/stz016}, \href
  {https://ui.adsabs.harvard.edu/abs/2019MNRAS.484..239D} {484, 239}

\bibitem[\protect\citeauthoryear{{Diemer} et~al.,}{{Diemer}
  et~al.}{2018}]{Diemer_2018}
{Diemer} B.,  et~al., 2018, \mn@doi [\apjs] {10.3847/1538-4365/aae387}, \href
  {https://ui.adsabs.harvard.edu/abs/2018ApJS..238...33D} {238, 33}

\bibitem[\protect\citeauthoryear{{Diemer} et~al.,}{{Diemer}
  et~al.}{2019}]{Diemer_2019}
{Diemer} B.,  et~al., 2019, \mn@doi [\mnras] {10.1093/mnras/stz1323}, \href
  {https://ui.adsabs.harvard.edu/abs/2019MNRAS.487.1529D} {487, 1529}

\bibitem[\protect\citeauthoryear{Dietterich}{Dietterich}{2000}]{Dietterich_ensembles}
Dietterich T.~G.,  2000, in Multiple Classifier Systems. Springer Berlin
  Heidelberg, Berlin, Heidelberg, pp 1--15

\bibitem[\protect\citeauthoryear{{Dillon} et~al.,}{{Dillon}
  et~al.}{2017}]{TensorFlow_probability}
{Dillon} J.~V.,  et~al., 2017, arXiv e-prints, \href
  {https://ui.adsabs.harvard.edu/abs/2017arXiv171110604D} {p. arXiv:1711.10604}

\bibitem[\protect\citeauthoryear{{Dolag}, {Borgani}, {Murante}  \&
  {Springel}}{{Dolag} et~al.}{2009}]{Dolag_2009}
{Dolag} K.,  {Borgani} S.,  {Murante} G.,   {Springel} V.,  2009, \mn@doi
  [\mnras] {10.1111/j.1365-2966.2009.15034.x}, \href
  {https://ui.adsabs.harvard.edu/abs/2009MNRAS.399..497D} {399, 497}

\bibitem[\protect\citeauthoryear{{Donnari} et~al.,}{{Donnari}
  et~al.}{2019}]{Donnari_2019}
{Donnari} M.,  et~al., 2019, \mn@doi [\mnras] {10.1093/mnras/stz712}, \href
  {https://ui.adsabs.harvard.edu/abs/2019MNRAS.485.4817D} {485, 4817}

\bibitem[\protect\citeauthoryear{{Duan}, {Avati}, {Ding}, {Thai}, {Basu}, {Ng}
  \& {Schuler}}{{Duan} et~al.}{2019}]{NGBoost}
{Duan} T.,  {Avati} A.,  {Ding} D.~Y.,  {Thai} K.~K.,  {Basu} S.,  {Ng} A.~Y.,
   {Schuler} A.,  2019, arXiv e-prints, \href
  {https://ui.adsabs.harvard.edu/abs/2019arXiv191003225D} {p. arXiv:1910.03225}

\bibitem[\protect\citeauthoryear{{Dubois} \& {Teyssier}}{{Dubois} \&
  {Teyssier}}{2008}]{Dubois_2008}
{Dubois} Y.,  {Teyssier} R.,  2008, \mn@doi [\aap]
  {10.1051/0004-6361:20078326}, \href
  {https://ui.adsabs.harvard.edu/abs/2008A&A...477...79D} {477, 79}

\bibitem[\protect\citeauthoryear{{Dubois}, {Pichon}, {Devriendt}, {Silk},
  {Haehnelt}, {Kimm}  \& {Slyz}}{{Dubois} et~al.}{2013}]{Dubois_2013}
{Dubois} Y.,  {Pichon} C.,  {Devriendt} J.,  {Silk} J.,  {Haehnelt} M.,  {Kimm}
  T.,   {Slyz} A.,  2013, \mn@doi [\mnras] {10.1093/mnras/sts224}, \href
  {https://ui.adsabs.harvard.edu/abs/2013MNRAS.428.2885D} {428, 2885}

\bibitem[\protect\citeauthoryear{{Dubois} et~al.,}{{Dubois}
  et~al.}{2014}]{Dubois_2014}
{Dubois} Y.,  et~al., 2014, \mn@doi [\mnras] {10.1093/mnras/stu1227}, \href
  {https://ui.adsabs.harvard.edu/abs/2014MNRAS.444.1453D} {444, 1453}

\bibitem[\protect\citeauthoryear{{Dutta}, {Khandai}  \& {Rana}}{{Dutta}
  et~al.}{2021}]{Dutta2021}
{Dutta} S.,  {Khandai} N.,   {Rana} S.,  2021, \mn@doi [\mnras]
  {10.1093/mnras/stab3618}, \href
  {https://ui.adsabs.harvard.edu/abs/2021MNRAS.tmp.3283D} {}

\bibitem[\protect\citeauthoryear{{Eisert}, {Pillepich}, {Nelson}, {Klessen},
  {Huertas-Company}  \& {Rodriguez-Gomez}}{{Eisert}
  et~al.}{2022}]{Eisert2022ErgoML}
{Eisert} L.,  {Pillepich} A.,  {Nelson} D.,  {Klessen} R.~S.,
  {Huertas-Company} M.,   {Rodriguez-Gomez} V.,  2022, arXiv e-prints, \href
  {https://ui.adsabs.harvard.edu/abs/2022arXiv220206967E} {p. arXiv:2202.06967}

\bibitem[\protect\citeauthoryear{F.R.S.}{F.R.S.}{1901}]{pearson_PCA}
F.R.S. K.~P.,  1901, \mn@doi [The London, Edinburgh, and Dublin Philosophical
  Magazine and Journal of Science] {10.1080/14786440109462720}, 2, 559

\bibitem[\protect\citeauthoryear{{Fall} \& {Efstathiou}}{{Fall} \&
  {Efstathiou}}{1980}]{Fall_Efstathiou_1980}
{Fall} S.~M.,  {Efstathiou} G.,  1980, \mn@doi [\mnras]
  {10.1093/mnras/193.2.189}, \href
  {https://ui.adsabs.harvard.edu/abs/1980MNRAS.193..189F} {193, 189}

\bibitem[\protect\citeauthoryear{Geurts, Ernst  \& Wehenkel}{Geurts
  et~al.}{2006}]{Geurts_2006_ExtraTrees}
Geurts P.,  Ernst D.,   Wehenkel L.,  2006, Machine Learning, 36, 3

\bibitem[\protect\citeauthoryear{{Gnedin}, {Weinberg}, {Pizagno}, {Prada}  \&
  {Rix}}{{Gnedin} et~al.}{2007}]{Gnedin_2007}
{Gnedin} O.~Y.,  {Weinberg} D.~H.,  {Pizagno} J.,  {Prada} F.,   {Rix} H.-W.,
  2007, \mn@doi [\apj] {10.1086/523256}, \href
  {https://ui.adsabs.harvard.edu/abs/2007ApJ...671.1115G} {671, 1115}

\bibitem[\protect\citeauthoryear{{Goodfellow}, {Shlens}  \&
  {Szegedy}}{{Goodfellow} et~al.}{2014}]{Goodfellow_adversarial_examples}
{Goodfellow} I.~J.,  {Shlens} J.,   {Szegedy} C.,  2014, arXiv e-prints, \href
  {https://ui.adsabs.harvard.edu/abs/2014arXiv1412.6572G} {p. arXiv:1412.6572}

\bibitem[\protect\citeauthoryear{{Gunn} \& {Gott}}{{Gunn} \&
  {Gott}}{1972}]{Gunn_1972}
{Gunn} J.~E.,  {Gott} J.~Richard I.,  1972, \mn@doi [\apj] {10.1086/151605},
  \href {https://ui.adsabs.harvard.edu/abs/1972ApJ...176....1G} {176, 1}

\bibitem[\protect\citeauthoryear{{Guo}, {Li}, {Zheng}, {Mo}, {Jing}, {Zu},
  {Lim}  \& {Xu}}{{Guo} et~al.}{2017}]{Guo}
{Guo} H.,  {Li} C.,  {Zheng} Z.,  {Mo} H.~J.,  {Jing} Y.~P.,  {Zu} Y.,  {Lim}
  S.~H.,   {Xu} H.,  2017, \mn@doi [\apj] {10.3847/1538-4357/aa85e7}, \href
  {https://ui.adsabs.harvard.edu/abs/2017ApJ...846...61G} {846, 61}

\bibitem[\protect\citeauthoryear{{Hadzhiyska}, {Bose}, {Eisenstein}  \&
  {Hernquist}}{{Hadzhiyska} et~al.}{2021}]{Hadzhiyska_HOD}
{Hadzhiyska} B.,  {Bose} S.,  {Eisenstein} D.,   {Hernquist} L.,  2021, \mn@doi
  [\mnras] {10.1093/mnras/staa3776}, \href
  {https://ui.adsabs.harvard.edu/abs/2021MNRAS.501.1603H} {501, 1603}

\bibitem[\protect\citeauthoryear{{Hearin}, {Watson}, {Becker}, {Reyes},
  {Berlind}  \& {Zentner}}{{Hearin} et~al.}{2014}]{CAM}
{Hearin} A.~P.,  {Watson} D.~F.,  {Becker} M.~R.,  {Reyes} R.,  {Berlind}
  A.~A.,   {Zentner} A.~R.,  2014, \mn@doi [\mnras] {10.1093/mnras/stu1443},
  \href {https://ui.adsabs.harvard.edu/abs/2014MNRAS.444..729H} {444, 729}

\bibitem[\protect\citeauthoryear{{Hearin}, {Zentner}, {van den Bosch},
  {Campbell}  \& {Tollerud}}{{Hearin} et~al.}{2016}]{Decorated_HOD}
{Hearin} A.~P.,  {Zentner} A.~R.,  {van den Bosch} F.~C.,  {Campbell} D.,
  {Tollerud} E.,  2016, \mn@doi [\mnras] {10.1093/mnras/stw840}, \href
  {https://ui.adsabs.harvard.edu/abs/2016MNRAS.460.2552H} {460, 2552}

\bibitem[\protect\citeauthoryear{{Ho}, {Farahi}, {Rau}  \& {Trac}}{{Ho}
  et~al.}{2021}]{Ho2021MLUncertainties}
{Ho} M.,  {Farahi} A.,  {Rau} M.~M.,   {Trac} H.,  2021, \mn@doi [\apj]
  {10.3847/1538-4357/abd101}, \href
  {https://ui.adsabs.harvard.edu/abs/2021ApJ...908..204H} {908, 204}

\bibitem[\protect\citeauthoryear{Hornik, Stinchcombe  \& White}{Hornik
  et~al.}{1989}]{Hornik_universal_approximator}
Hornik K.,  Stinchcombe M.,   White H.,  1989, \mn@doi [Neural Networks]
  {https://doi.org/10.1016/0893-6080(89)90020-8}, 2, 359

\bibitem[\protect\citeauthoryear{Hotelling}{Hotelling}{1936}]{hotelling_PCA}
Hotelling H.,  1936, Biometrika, 28, 321

\bibitem[\protect\citeauthoryear{{Jeffrey} \& {Wandelt}}{{Jeffrey} \&
  {Wandelt}}{2020}]{Jeffrey_2020}
{Jeffrey} N.,  {Wandelt} B.~D.,  2020, arXiv e-prints, \href
  {https://ui.adsabs.harvard.edu/abs/2020arXiv201105991J} {p. arXiv:2011.05991}

\bibitem[\protect\citeauthoryear{{Jiang} et~al.,}{{Jiang}
  et~al.}{2019}]{Jiang_2019}
{Jiang} F.,  et~al., 2019, \mn@doi [\mnras] {10.1093/mnras/stz1952}, \href
  {https://ui.adsabs.harvard.edu/abs/2019MNRAS.488.4801J} {488, 4801}

\bibitem[\protect\citeauthoryear{{Jo} \& {Kim}}{{Jo} \&
  {Kim}}{2019}]{Yongseok_2019}
{Jo} Y.,  {Kim} J.-h.,  2019, \mn@doi [\mnras] {10.1093/mnras/stz2304}, \href
  {https://ui.adsabs.harvard.edu/abs/2019MNRAS.489.3565J} {489, 3565}

\bibitem[\protect\citeauthoryear{{Kamdar}, {Turk}  \& {Brunner}}{{Kamdar}
  et~al.}{2016}]{Kamdar_2016}
{Kamdar} H.~M.,  {Turk} M.~J.,   {Brunner} R.~J.,  2016, \mn@doi [\mnras]
  {10.1093/mnras/stv2981}, \href
  {https://ui.adsabs.harvard.edu/abs/2016MNRAS.457.1162K} {457, 1162}

\bibitem[\protect\citeauthoryear{{Kasmanoff}, {Villaescusa-Navarro}, {Tinker}
  \& {Ho}}{{Kasmanoff} et~al.}{2020}]{Kasmanoff_2020}
{Kasmanoff} N.,  {Villaescusa-Navarro} F.,  {Tinker} J.,   {Ho} S.,  2020,
  arXiv e-prints, \href {https://ui.adsabs.harvard.edu/abs/2020arXiv201200186K}
  {p. arXiv:2012.00186}

\bibitem[\protect\citeauthoryear{{Kaviraj} et~al.,}{{Kaviraj}
  et~al.}{2017}]{Kaviraj_2017}
{Kaviraj} S.,  et~al., 2017, \mn@doi [\mnras] {10.1093/mnras/stx126}, \href
  {https://ui.adsabs.harvard.edu/abs/2017MNRAS.467.4739K} {467, 4739}

\bibitem[\protect\citeauthoryear{Kimm}{Kimm}{2012}]{taysun2012a}
Kimm T.,  2012, PhD thesis, Oxford University, UK

\bibitem[\protect\citeauthoryear{{Kingma} \& {Ba}}{{Kingma} \&
  {Ba}}{2014}]{Adam_Kingma}
{Kingma} D.~P.,  {Ba} J.,  2014, arXiv e-prints, \href
  {https://ui.adsabs.harvard.edu/abs/2014arXiv1412.6980K} {p. arXiv:1412.6980}

\bibitem[\protect\citeauthoryear{{Klambauer}, {Unterthiner}, {Mayr}  \&
  {Hochreiter}}{{Klambauer} et~al.}{2017}]{Klambauer_SELU}
{Klambauer} G.,  {Unterthiner} T.,  {Mayr} A.,   {Hochreiter} S.,  2017, arXiv
  e-prints, \href {https://ui.adsabs.harvard.edu/abs/2017arXiv170602515K} {p.
  arXiv:1706.02515}

\bibitem[\protect\citeauthoryear{{Kodi Ramanah}, {Wojtak}, {Ansari}, {Gall}  \&
  {Hjorth}}{{Kodi Ramanah} et~al.}{2020}]{Ramanah2020MLUncertainties}
{Kodi Ramanah} D.,  {Wojtak} R.,  {Ansari} Z.,  {Gall} C.,   {Hjorth} J.,
  2020, \mn@doi [\mnras] {10.1093/mnras/staa2886}, \href
  {https://ui.adsabs.harvard.edu/abs/2020MNRAS.499.1985K} {499, 1985}

\bibitem[\protect\citeauthoryear{{Kodi Ramanah}, {Wojtak}  \& {Arendse}}{{Kodi
  Ramanah} et~al.}{2021}]{Ramanah2021MLUncertainties}
{Kodi Ramanah} D.,  {Wojtak} R.,   {Arendse} N.,  2021, \mn@doi [\mnras]
  {10.1093/mnras/staa3922}, \href
  {https://ui.adsabs.harvard.edu/abs/2021MNRAS.501.4080K} {501, 4080}

\bibitem[\protect\citeauthoryear{{Komatsu} et~al.,}{{Komatsu}
  et~al.}{2011}]{Komatsu_2011}
{Komatsu} E.,  et~al., 2011, \mn@doi [\apjs] {10.1088/0067-0049/192/2/18},
  \href {https://ui.adsabs.harvard.edu/abs/2011ApJS..192...18K} {192, 18}

\bibitem[\protect\citeauthoryear{{Kravtsov}}{{Kravtsov}}{2013}]{Kravtsov_2013}
{Kravtsov} A.~V.,  2013, \mn@doi [\apjl] {10.1088/2041-8205/764/2/L31}, \href
  {https://ui.adsabs.harvard.edu/abs/2013ApJ...764L..31K} {764, L31}

\bibitem[\protect\citeauthoryear{{Kravtsov}, {Berlind}, {Wechsler}, {Klypin},
  {Gottl{\"o}ber}, {Allgood}  \& {Primack}}{{Kravtsov}
  et~al.}{2004}]{Kravtsov_2004}
{Kravtsov} A.~V.,  {Berlind} A.~A.,  {Wechsler} R.~H.,  {Klypin} A.~A.,
  {Gottl{\"o}ber} S.,  {Allgood} B.,   {Primack} J.~R.,  2004, \mn@doi [\apj]
  {10.1086/420959}, \href
  {https://ui.adsabs.harvard.edu/abs/2004ApJ...609...35K} {609, 35}

\bibitem[\protect\citeauthoryear{{Laigle} et~al.,}{{Laigle}
  et~al.}{2019}]{Laigle_2019}
{Laigle} C.,  et~al., 2019, \mn@doi [\mnras] {10.1093/mnras/stz1054}, \href
  {https://ui.adsabs.harvard.edu/abs/2019MNRAS.486.5104L} {486, 5104}

\bibitem[\protect\citeauthoryear{{Lakshminarayanan}, {Pritzel}  \&
  {Blundell}}{{Lakshminarayanan}
  et~al.}{2016}]{Lakshminarayanan_deep_ensembles}
{Lakshminarayanan} B.,  {Pritzel} A.,   {Blundell} C.,  2016, arXiv e-prints,
  \href {https://ui.adsabs.harvard.edu/abs/2016arXiv161201474L} {p.
  arXiv:1612.01474}

\bibitem[\protect\citeauthoryear{LeCun, Bottou, Orr  \& Muller}{LeCun
  et~al.}{1998}]{lecun-98b}
LeCun Y.,  Bottou L.,  Orr G.,   Muller K.,  1998, in Orr G.,  K. M.,  eds,
  Neural Networks: Tricks of the trade. Springer

\bibitem[\protect\citeauthoryear{{Lehmann}, {Mao}, {Becker}, {Skillman}  \&
  {Wechsler}}{{Lehmann} et~al.}{2017}]{Lehmann2017}
{Lehmann} B.~V.,  {Mao} Y.-Y.,  {Becker} M.~R.,  {Skillman} S.~W.,   {Wechsler}
  R.~H.,  2017, \mn@doi [\apj] {10.3847/1538-4357/834/1/37}, \href
  {https://ui.adsabs.harvard.edu/abs/2017ApJ...834...37L} {834, 37}

\bibitem[\protect\citeauthoryear{{Li}, {McGaugh}, {Lelli}, {Tian}, {Schombert}
  \& {Ko}}{{Li} et~al.}{2022}]{Li_2022}
{Li} P.,  {McGaugh} S.~S.,  {Lelli} F.,  {Tian} Y.,  {Schombert} J.~M.,   {Ko}
  C.-M.,  2022, arXiv e-prints, \href
  {https://ui.adsabs.harvard.edu/abs/2022arXiv220203421L} {p. arXiv:2202.03421}

\bibitem[\protect\citeauthoryear{{Lim}, {Barnes}, {Vogelsberger}, {Mo},
  {Nelson}, {Pillepich}, {Dolag}  \& {Marinacci}}{{Lim}
  et~al.}{2021}]{Lim2021Gas}
{Lim} S.~H.,  {Barnes} D.,  {Vogelsberger} M.,  {Mo} H.~J.,  {Nelson} D.,
  {Pillepich} A.,  {Dolag} K.,   {Marinacci} F.,  2021, \mn@doi [\mnras]
  {10.1093/mnras/stab1172}, \href
  {https://ui.adsabs.harvard.edu/abs/2021MNRAS.504.5131L} {504, 5131}

\bibitem[\protect\citeauthoryear{{Lovell}, {Wilkins}, {Thomas}, {Schaller},
  {Baugh}, {Fabbian}  \& {Bah{\'e}}}{{Lovell} et~al.}{2022}]{Lovell_2021}
{Lovell} C.~C.,  {Wilkins} S.~M.,  {Thomas} P.~A.,  {Schaller} M.,  {Baugh}
  C.~M.,  {Fabbian} G.,   {Bah{\'e}} Y.,  2022, \mn@doi [\mnras]
  {10.1093/mnras/stab3221}, \href
  {https://ui.adsabs.harvard.edu/abs/2022MNRAS.509.5046L} {509, 5046}

\bibitem[\protect\citeauthoryear{{Machado Poletti Valle}, {Avestruz}, {Barnes},
  {Farahi}, {Lau}  \& {Nagai}}{{Machado Poletti Valle}
  et~al.}{2021}]{Machado2021SHAP}
{Machado Poletti Valle} L.~F.,  {Avestruz} C.,  {Barnes} D.~J.,  {Farahi} A.,
  {Lau} E.~T.,   {Nagai} D.,  2021, \mn@doi [\mnras] {10.1093/mnras/stab2252},
  \href {https://ui.adsabs.harvard.edu/abs/2021MNRAS.tmp.2050M} {}

\bibitem[\protect\citeauthoryear{{Man}, {Peng}, {Shi}, {Kong}, {Zhang}, {Dou}
  \& {Guo}}{{Man} et~al.}{2019}]{Man_2019}
{Man} Z.-Y.,  {Peng} Y.-J.,  {Shi} J.-J.,  {Kong} X.,  {Zhang} C.-P.,  {Dou}
  J.,   {Guo} K.-X.,  2019, \mn@doi [\apj] {10.3847/1538-4357/ab2ece}, \href
  {https://ui.adsabs.harvard.edu/abs/2019ApJ...881...74M} {881, 74}

\bibitem[\protect\citeauthoryear{{Mancera Pi{\~n}a}, {Posti}, {Fraternali},
  {Adams}  \& {Oosterloo}}{{Mancera Pi{\~n}a} et~al.}{2021}]{Pina}
{Mancera Pi{\~n}a} P.~E.,  {Posti} L.,  {Fraternali} F.,  {Adams} E. A.~K.,
  {Oosterloo} T.,  2021, \mn@doi [\aap] {10.1051/0004-6361/202039340}, \href
  {https://ui.adsabs.harvard.edu/abs/2021A&A...647A..76M} {647, A76}

\bibitem[\protect\citeauthoryear{{Marinacci} et~al.,}{{Marinacci}
  et~al.}{2018}]{Marinacci_2018}
{Marinacci} F.,  et~al., 2018, \mn@doi [\mnras] {10.1093/mnras/sty2206}, \href
  {https://ui.adsabs.harvard.edu/abs/2018MNRAS.480.5113M} {480, 5113}

\bibitem[\protect\citeauthoryear{{McCarthy}, {Schaye}, {Bird}  \& {Le
  Brun}}{{McCarthy} et~al.}{2017}]{McCarthy2017Bahamas}
{McCarthy} I.~G.,  {Schaye} J.,  {Bird} S.,   {Le Brun} A. M.~C.,  2017,
  \mn@doi [\mnras] {10.1093/mnras/stw2792}, \href
  {https://ui.adsabs.harvard.edu/abs/2017MNRAS.465.2936M} {465, 2936}

\bibitem[\protect\citeauthoryear{{McGibbon} \& {Khochfar}}{{McGibbon} \&
  {Khochfar}}{2021}]{McGibbon}
{McGibbon} R.,  {Khochfar} S.,  2021, arXiv e-prints, \href
  {https://ui.adsabs.harvard.edu/abs/2021arXiv211208424M} {p. arXiv:2112.08424}

\bibitem[\protect\citeauthoryear{{Mo}, {Mao}  \& {White}}{{Mo}
  et~al.}{1998}]{MMW_1998}
{Mo} H.~J.,  {Mao} S.,   {White} S. D.~M.,  1998, \mn@doi [\mnras]
  {10.1046/j.1365-8711.1998.01227.x}, \href
  {https://ui.adsabs.harvard.edu/abs/1998MNRAS.295..319M} {295, 319}

\bibitem[\protect\citeauthoryear{{Moster}, {Somerville}, {Maulbetsch}, {van den
  Bosch}, {Macci{\`o}}, {Naab}  \& {Oser}}{{Moster} et~al.}{2010}]{Moster2010}
{Moster} B.~P.,  {Somerville} R.~S.,  {Maulbetsch} C.,  {van den Bosch} F.~C.,
  {Macci{\`o}} A.~V.,  {Naab} T.,   {Oser} L.,  2010, \mn@doi [\apj]
  {10.1088/0004-637X/710/2/903}, \href
  {https://ui.adsabs.harvard.edu/abs/2010ApJ...710..903M} {710, 903}

\bibitem[\protect\citeauthoryear{{Moster}, {Naab}, {Lindstr{\"o}m}  \&
  {O'Leary}}{{Moster} et~al.}{2021}]{Moster_GalaxyNet}
{Moster} B.~P.,  {Naab} T.,  {Lindstr{\"o}m} M.,   {O'Leary} J.~A.,  2021,
  \mn@doi [\mnras] {10.1093/mnras/stab1449}, \href
  {https://ui.adsabs.harvard.edu/abs/2021MNRAS.tmp.1690M} {}

\bibitem[\protect\citeauthoryear{{Naiman} et~al.,}{{Naiman}
  et~al.}{2018}]{Naiman_2018}
{Naiman} J.~P.,  et~al., 2018, \mn@doi [\mnras] {10.1093/mnras/sty618}, \href
  {https://ui.adsabs.harvard.edu/abs/2018MNRAS.477.1206N} {477, 1206}

\bibitem[\protect\citeauthoryear{{Navarro}, {Frenk}  \& {White}}{{Navarro}
  et~al.}{1996}]{Navarro_1996}
{Navarro} J.~F.,  {Frenk} C.~S.,   {White} S. D.~M.,  1996, \mn@doi [\apj]
  {10.1086/177173}, \href
  {https://ui.adsabs.harvard.edu/abs/1996ApJ...462..563N} {462, 563}

\bibitem[\protect\citeauthoryear{{Nelson} et~al.,}{{Nelson}
  et~al.}{2018}]{Nelson_2018}
{Nelson} D.,  et~al., 2018, \mn@doi [\mnras] {10.1093/mnras/stx3040}, \href
  {https://ui.adsabs.harvard.edu/abs/2018MNRAS.475..624N} {475, 624}

\bibitem[\protect\citeauthoryear{{Nelson} et~al.,}{{Nelson}
  et~al.}{2019}]{Nelson_2019}
{Nelson} D.,  et~al., 2019, \mn@doi [Computational Astrophysics and Cosmology]
  {10.1186/s40668-019-0028-x}, \href
  {https://ui.adsabs.harvard.edu/abs/2019ComAC...6....2N} {6, 2}

\bibitem[\protect\citeauthoryear{Nix \& Weigend}{Nix \&
  Weigend}{1994}]{Nix_mean_variance}
Nix D.,  Weigend A.,  1994, in Proceedings of 1994 IEEE International
  Conference on Neural Networks (ICNN'94). pp 55--60 vol.1,
  \mn@doi{10.1109/ICNN.1994.374138}

\bibitem[\protect\citeauthoryear{{Ntampaka} et~al.,}{{Ntampaka}
  et~al.}{2019}]{Ntampaka2019MLDeepDream}
{Ntampaka} M.,  et~al., 2019, \mn@doi [\apj] {10.3847/1538-4357/ab14eb}, \href
  {https://ui.adsabs.harvard.edu/abs/2019ApJ...876...82N} {876, 82}

\bibitem[\protect\citeauthoryear{{Olson}, {Bartley}, {Urbanowicz}  \&
  {Moore}}{{Olson} et~al.}{2016}]{Olson_2016}
{Olson} R.~S.,  {Bartley} N.,  {Urbanowicz} R.~J.,   {Moore} J.~H.,  2016,
  arXiv e-prints, \href {https://ui.adsabs.harvard.edu/abs/2016arXiv160306212O}
  {p. arXiv:1603.06212}

\bibitem[\protect\citeauthoryear{{Paranjape}, {Kova{\v{c}}}, {Hartley}  \&
  {Pahwa}}{{Paranjape} et~al.}{2015}]{Paranjape_HOD}
{Paranjape} A.,  {Kova{\v{c}}} K.,  {Hartley} W.~G.,   {Pahwa} I.,  2015,
  \mn@doi [\mnras] {10.1093/mnras/stv2137}, \href
  {https://ui.adsabs.harvard.edu/abs/2015MNRAS.454.3030P} {454, 3030}

\bibitem[\protect\citeauthoryear{Pedregosa et~al.}{Pedregosa
  et~al.}{2011}]{scikit-learn}
Pedregosa F.,  et~al., 2011, Journal of Machine Learning Research, 12, 2825

\bibitem[\protect\citeauthoryear{{Pillepich} et~al.,}{{Pillepich}
  et~al.}{2018a}]{Pillepich_2018_methods}
{Pillepich} A.,  et~al., 2018a, \mn@doi [\mnras] {10.1093/mnras/stx2656}, \href
  {https://ui.adsabs.harvard.edu/abs/2018MNRAS.473.4077P} {473, 4077}

\bibitem[\protect\citeauthoryear{{Pillepich} et~al.,}{{Pillepich}
  et~al.}{2018b}]{Pillepich_2018_results}
{Pillepich} A.,  et~al., 2018b, \mn@doi [\mnras] {10.1093/mnras/stx3112}, \href
  {https://ui.adsabs.harvard.edu/abs/2018MNRAS.475..648P} {475, 648}

\bibitem[\protect\citeauthoryear{{Pillepich} et~al.,}{{Pillepich}
  et~al.}{2019}]{Pillepich_2019}
{Pillepich} A.,  et~al., 2019, \mn@doi [\mnras] {10.1093/mnras/stz2338}, \href
  {https://ui.adsabs.harvard.edu/abs/2019MNRAS.490.3196P} {490, 3196}

\bibitem[\protect\citeauthoryear{{Planck Collaboration}}{{Planck
  Collaboration}}{2016}]{Planck_Parameters_2016}
{Planck Collaboration} 2016, \mn@doi [\aap] {10.1051/0004-6361/201525830},
  \href {https://ui.adsabs.harvard.edu/abs/2016A&A...594A..13P} {594, A13}

\bibitem[\protect\citeauthoryear{{Posti}, {Pezzulli}, {Fraternali}  \& {Di
  Teodoro}}{{Posti} et~al.}{2018}]{Posti}
{Posti} L.,  {Pezzulli} G.,  {Fraternali} F.,   {Di Teodoro} E.~M.,  2018,
  \mn@doi [\mnras] {10.1093/mnras/stx3168}, \href
  {https://ui.adsabs.harvard.edu/abs/2018MNRAS.475..232P} {475, 232}

\bibitem[\protect\citeauthoryear{{Power}, {Navarro}, {Jenkins}, {Frenk},
  {White}, {Springel}, {Stadel}  \& {Quinn}}{{Power} et~al.}{2003}]{Power_2003}
{Power} C.,  {Navarro} J.~F.,  {Jenkins} A.,  {Frenk} C.~S.,  {White} S.~D.~M.,
   {Springel} V.,  {Stadel} J.,   {Quinn} T.,  2003, \mn@doi [\mnras]
  {10.1046/j.1365-8711.2003.05925.x}, \href
  {https://ui.adsabs.harvard.edu/abs/2003MNRAS.338...14P} {338, 14}

\bibitem[\protect\citeauthoryear{{Ragagnin}, {Fumagalli}, {Castro}, {Dolag},
  {Saro}, {Costanzi}  \& {Bocquet}}{{Ragagnin} et~al.}{2021}]{Ragagnin2021HOD}
{Ragagnin} A.,  {Fumagalli} A.,  {Castro} T.,  {Dolag} K.,  {Saro} A.,
  {Costanzi} M.,   {Bocquet} S.,  2021, arXiv e-prints, \href
  {https://ui.adsabs.harvard.edu/abs/2021arXiv211005498R} {p. arXiv:2110.05498}

\bibitem[\protect\citeauthoryear{{Rasera} \& {Teyssier}}{{Rasera} \&
  {Teyssier}}{2006}]{Rasera_2006}
{Rasera} Y.,  {Teyssier} R.,  2006, \mn@doi [\aap]
  {10.1051/0004-6361:20053116}, \href
  {https://ui.adsabs.harvard.edu/abs/2006A&A...445....1R} {445, 1}

\bibitem[\protect\citeauthoryear{{Reddick}, {Wechsler}, {Tinker}  \&
  {Behroozi}}{{Reddick} et~al.}{2013}]{Reddick2013}
{Reddick} R.~M.,  {Wechsler} R.~H.,  {Tinker} J.~L.,   {Behroozi} P.~S.,  2013,
  \mn@doi [\apj] {10.1088/0004-637X/771/1/30}, \href
  {https://ui.adsabs.harvard.edu/abs/2013ApJ...771...30R} {771, 30}

\bibitem[\protect\citeauthoryear{{Reddick}, {Tinker}, {Wechsler}  \&
  {Lu}}{{Reddick} et~al.}{2014}]{Reddick_2014}
{Reddick} R.~M.,  {Tinker} J.~L.,  {Wechsler} R.~H.,   {Lu} Y.,  2014, \mn@doi
  [\apj] {10.1088/0004-637X/783/2/118}, \href
  {https://ui.adsabs.harvard.edu/abs/2014ApJ...783..118R} {783, 118}

\bibitem[\protect\citeauthoryear{{Rohr} et~al.,}{{Rohr} et~al.}{2022}]{Rohr}
{Rohr} E.,  et~al., 2022, \mn@doi [\mnras] {10.1093/mnras/stab3625}, \href
  {https://ui.adsabs.harvard.edu/abs/2022MNRAS.510.3967R} {510, 3967}

\bibitem[\protect\citeauthoryear{{Rumelhart}, {Hinton}  \&
  {Williams}}{{Rumelhart} et~al.}{1986}]{Rumelhart_backprop}
{Rumelhart} D.~E.,  {Hinton} G.~E.,   {Williams} R.~J.,  1986, \mn@doi [\nat]
  {10.1038/323533a0}, \href
  {https://ui.adsabs.harvard.edu/abs/1986Natur.323..533R} {323, 533}

\bibitem[\protect\citeauthoryear{{Sales}, {Navarro}, {Schaye}, {Dalla Vecchia},
  {Springel}, {Haas}  \& {Helmi}}{{Sales} et~al.}{2009}]{Sales_2009}
{Sales} L.~V.,  {Navarro} J.~F.,  {Schaye} J.,  {Dalla Vecchia} C.,  {Springel}
  V.,  {Haas} M.~R.,   {Helmi} A.,  2009, \mn@doi [\mnras]
  {10.1111/j.1745-3933.2009.00725.x}, \href
  {https://ui.adsabs.harvard.edu/abs/2009MNRAS.399L..64S} {399, L64}

\bibitem[\protect\citeauthoryear{{Sales}, {Navarro}, {Theuns}, {Schaye},
  {White}, {Frenk}, {Crain}  \& {Dalla Vecchia}}{{Sales}
  et~al.}{2012}]{Sales_2012}
{Sales} L.~V.,  {Navarro} J.~F.,  {Theuns} T.,  {Schaye} J.,  {White} S. D.~M.,
   {Frenk} C.~S.,  {Crain} R.~A.,   {Dalla Vecchia} C.,  2012, \mn@doi [\mnras]
  {10.1111/j.1365-2966.2012.20975.x}, \href
  {https://ui.adsabs.harvard.edu/abs/2012MNRAS.423.1544S} {423, 1544}

\bibitem[\protect\citeauthoryear{{Shao} et~al.,}{{Shao}
  et~al.}{2021}]{Shao_2021}
{Shao} H.,  et~al., 2021, arXiv e-prints, \href
  {https://ui.adsabs.harvard.edu/abs/2021arXiv210904484S} {p. arXiv:2109.04484}

\bibitem[\protect\citeauthoryear{{Springel}}{{Springel}}{2010}]{Springel_2010}
{Springel} V.,  2010, \mn@doi [\mnras] {10.1111/j.1365-2966.2009.15715.x},
  \href {https://ui.adsabs.harvard.edu/abs/2010MNRAS.401..791S} {401, 791}

\bibitem[\protect\citeauthoryear{{Springel} \& {Hernquist}}{{Springel} \&
  {Hernquist}}{2003}]{Springel_2003}
{Springel} V.,  {Hernquist} L.,  2003, \mn@doi [\mnras]
  {10.1046/j.1365-8711.2003.06206.x}, \href
  {https://ui.adsabs.harvard.edu/abs/2003MNRAS.339..289S} {339, 289}

\bibitem[\protect\citeauthoryear{{Springel}, {White}, {Tormen}  \&
  {Kauffmann}}{{Springel} et~al.}{2001}]{Springel_2001}
{Springel} V.,  {White} S. D.~M.,  {Tormen} G.,   {Kauffmann} G.,  2001,
  \mn@doi [\mnras] {10.1046/j.1365-8711.2001.04912.x}, \href
  {https://ui.adsabs.harvard.edu/abs/2001MNRAS.328..726S} {328, 726}

\bibitem[\protect\citeauthoryear{{Springel} et~al.,}{{Springel}
  et~al.}{2018}]{Springel_2018}
{Springel} V.,  et~al., 2018, \mn@doi [\mnras] {10.1093/mnras/stx3304}, \href
  {https://ui.adsabs.harvard.edu/abs/2018MNRAS.475..676S} {475, 676}

\bibitem[\protect\citeauthoryear{{Stevens}, {Lagos}, {Contreras}, {Croton},
  {Padilla}, {Schaller}, {Schaye}  \& {Theuns}}{{Stevens}
  et~al.}{2017}]{Stevens}
{Stevens} A. R.~H.,  {Lagos} C. d.~P.,  {Contreras} S.,  {Croton} D.~J.,
  {Padilla} N.~D.,  {Schaller} M.,  {Schaye} J.,   {Theuns} T.,  2017, \mn@doi
  [\mnras] {10.1093/mnras/stx243}, \href
  {https://ui.adsabs.harvard.edu/abs/2017MNRAS.467.2066S} {467, 2066}

\bibitem[\protect\citeauthoryear{{Stiskalek}, {Desmond}, {Holvey}  \&
  {Jones}}{{Stiskalek} et~al.}{2021}]{Stiskalek2021}
{Stiskalek} R.,  {Desmond} H.,  {Holvey} T.,   {Jones} M.~G.,  2021, \mn@doi
  [\mnras] {10.1093/mnras/stab1845}, \href
  {https://ui.adsabs.harvard.edu/abs/2021MNRAS.506.3205S} {506, 3205}

\bibitem[\protect\citeauthoryear{{Teyssier}}{{Teyssier}}{2002}]{Teyssier_2002}
{Teyssier} R.,  2002, \mn@doi [\aap] {10.1051/0004-6361:20011817}, \href
  {https://ui.adsabs.harvard.edu/abs/2002A&A...385..337T} {385, 337}

\bibitem[\protect\citeauthoryear{{Trujillo-Gomez}, {Klypin}, {Primack}  \&
  {Romanowsky}}{{Trujillo-Gomez} et~al.}{2011}]{Trujillo}
{Trujillo-Gomez} S.,  {Klypin} A.,  {Primack} J.,   {Romanowsky} A.~J.,  2011,
  \mn@doi [\apj] {10.1088/0004-637X/742/1/16}, \href
  {https://ui.adsabs.harvard.edu/abs/2011ApJ...742...16T} {742, 16}

\bibitem[\protect\citeauthoryear{{Tweed}, {Devriendt}, {Blaizot}, {Colombi}  \&
  {Slyz}}{{Tweed} et~al.}{2009}]{Tweed_2009}
{Tweed} D.,  {Devriendt} J.,  {Blaizot} J.,  {Colombi} S.,   {Slyz} A.,  2009,
  \mn@doi [\aap] {10.1051/0004-6361/200911787}, \href
  {https://ui.adsabs.harvard.edu/abs/2009A&A...506..647T} {506, 647}

\bibitem[\protect\citeauthoryear{{Valentin Jospin}, {Buntine}, {Boussaid},
  {Laga}  \& {Bennamoun}}{{Valentin Jospin} et~al.}{2020}]{BayesNN}
{Valentin Jospin} L.,  {Buntine} W.,  {Boussaid} F.,  {Laga} H.,   {Bennamoun}
  M.,  2020, arXiv e-prints, \href
  {https://ui.adsabs.harvard.edu/abs/2020arXiv200706823V} {p. arXiv:2007.06823}

\bibitem[\protect\citeauthoryear{{Villaescusa-Navarro}
  et~al.,}{{Villaescusa-Navarro} et~al.}{2021a}]{Villaescusa_Navarro_2021}
{Villaescusa-Navarro} F.,  et~al., 2021a, arXiv e-prints, \href
  {https://ui.adsabs.harvard.edu/abs/2021arXiv210909747V} {p. arXiv:2109.09747}

\bibitem[\protect\citeauthoryear{{Villaescusa-Navarro}
  et~al.,}{{Villaescusa-Navarro} et~al.}{2021b}]{Villaescusa_Navarro_2021_marg}
{Villaescusa-Navarro} F.,  et~al., 2021b, arXiv e-prints, \href
  {https://ui.adsabs.harvard.edu/abs/2021arXiv210910360V} {p. arXiv:2109.10360}

\bibitem[\protect\citeauthoryear{{Villaescusa-Navarro}
  et~al.,}{{Villaescusa-Navarro} et~al.}{2021c}]{CAMELS_2021}
{Villaescusa-Navarro} F.,  et~al., 2021c, \mn@doi [\apj]
  {10.3847/1538-4357/abf7ba}, \href
  {https://ui.adsabs.harvard.edu/abs/2021ApJ...915...71V} {915, 71}

\bibitem[\protect\citeauthoryear{{Villaescusa-Navarro}
  et~al.,}{{Villaescusa-Navarro} et~al.}{2022a}]{CAMELS_2022}
{Villaescusa-Navarro} F.,  et~al., 2022a, arXiv e-prints, \href
  {https://ui.adsabs.harvard.edu/abs/2022arXiv220101300V} {p. arXiv:2201.01300}

\bibitem[\protect\citeauthoryear{{Villaescusa-Navarro}
  et~al.,}{{Villaescusa-Navarro} et~al.}{2022b}]{Villaescusa2022SingleGalaxy}
{Villaescusa-Navarro} F.,  et~al., 2022b, arXiv e-prints, \href
  {https://ui.adsabs.harvard.edu/abs/2022arXiv220102202V} {p. arXiv:2201.02202}

\bibitem[\protect\citeauthoryear{{Villanueva-Domingo}
  et~al.,}{{Villanueva-Domingo} et~al.}{2021}]{Villanueva_Domingo_2021}
{Villanueva-Domingo} P.,  et~al., 2021, arXiv e-prints, \href
  {https://ui.adsabs.harvard.edu/abs/2021arXiv211108683V} {p. arXiv:2111.08683}

\bibitem[\protect\citeauthoryear{{Wechsler}, {DeRose}, {Busha}, {Becker},
  {Rykoff}  \& {Evrard}}{{Wechsler} et~al.}{2021}]{Wechsler2021Addgals}
{Wechsler} R.~H.,  {DeRose} J.,  {Busha} M.~T.,  {Becker} M.~R.,  {Rykoff} E.,
   {Evrard} A.,  2021, arXiv e-prints, \href
  {https://ui.adsabs.harvard.edu/abs/2021arXiv210512105W} {p. arXiv:2105.12105}

\bibitem[\protect\citeauthoryear{{Weinberger} et~al.,}{{Weinberger}
  et~al.}{2017}]{Weinberger_2017}
{Weinberger} R.,  et~al., 2017, \mn@doi [\mnras] {10.1093/mnras/stw2944}, \href
  {https://ui.adsabs.harvard.edu/abs/2017MNRAS.465.3291W} {465, 3291}

\bibitem[\protect\citeauthoryear{{Yang}, {Gao}, {Frenk}, {Grand}, {Guo}, {Liao}
   \& {Shao}}{{Yang} et~al.}{2021}]{Yang_2021}
{Yang} H.,  {Gao} L.,  {Frenk} C.~S.,  {Grand} R. J.~J.,  {Guo} Q.,  {Liao} S.,
    {Shao} S.,  2021, arXiv e-prints, \href
  {https://ui.adsabs.harvard.edu/abs/2021arXiv211004434Y} {p. arXiv:2110.04434}

\bibitem[\protect\citeauthoryear{{Zemp}, {Gnedin}, {Gnedin}  \&
  {Kravtsov}}{{Zemp} et~al.}{2011}]{Zemp2011Shapes}
{Zemp} M.,  {Gnedin} O.~Y.,  {Gnedin} N.~Y.,   {Kravtsov} A.~V.,  2011, \mn@doi
  [\apjs] {10.1088/0067-0049/197/2/30}, \href
  {https://ui.adsabs.harvard.edu/abs/2011ApJS..197...30Z} {197, 30}

\bibitem[\protect\citeauthoryear{{Zhang}, {Wang}, {Zhang}, {Sun}, {He},
  {Contardo}, {Villaescusa-Navarro}  \& {Ho}}{{Zhang}
  et~al.}{2019}]{Zhang_2019}
{Zhang} X.,  {Wang} Y.,  {Zhang} W.,  {Sun} Y.,  {He} S.,  {Contardo} G.,
  {Villaescusa-Navarro} F.,   {Ho} S.,  2019, arXiv e-prints, \href
  {https://ui.adsabs.harvard.edu/abs/2019arXiv190205965Z} {p. arXiv:1902.05965}

\bibitem[\protect\citeauthoryear{{de Jong} \& {Lacey}}{{de Jong} \&
  {Lacey}}{2000}]{de_Jong_2000}
{de Jong} R.~S.,  {Lacey} C.,  2000, \mn@doi [\apj] {10.1086/317840}, \href
  {https://ui.adsabs.harvard.edu/abs/2000ApJ...545..781D} {545, 781}

\bibitem[\protect\citeauthoryear{{de Santi}, {Rodrigues}, {Montero-Dorta},
  {Abramo}, {Tucci}  \& {Artale}}{{de Santi} et~al.}{2022}]{de_Santi_2022}
{de Santi} N. S.~M.,  {Rodrigues} N. V.~N.,  {Montero-Dorta} A.~D.,  {Abramo}
  L.~R.,  {Tucci} B.,   {Artale} M.~C.,  2022, arXiv e-prints, \href
  {https://ui.adsabs.harvard.edu/abs/2022arXiv220106054D} {p. arXiv:2201.06054}

\bibitem[\protect\citeauthoryear{{de los Rios}, {Peta{\v{c}}}, {Zaldivar},
  {Bonaventura}, {Calore}  \& {Iocco}}{{de los Rios}
  et~al.}{2021}]{de_los_Rios_2021}
{de los Rios} M.~E.,  {Peta{\v{c}}} M.,  {Zaldivar} B.,  {Bonaventura} N.~R.,
  {Calore} F.,   {Iocco} F.,  2021, arXiv e-prints, \href
  {https://ui.adsabs.harvard.edu/abs/2021arXiv211108725D} {p. arXiv:2111.08725}

\bibitem[\protect\citeauthoryear{{von Marttens} et~al.,}{{von Marttens}
  et~al.}{2021}]{vonMarttens_2021}
{von Marttens} R.,  et~al., 2021, arXiv e-prints, \href
  {https://ui.adsabs.harvard.edu/abs/2021arXiv211101185V} {p. arXiv:2111.01185}

\makeatother
\end{thebibliography}

\label{lastpage}
\end{document}